%% file: Articulo.tex
\documentclass[preprints,article,accept,moreauthors]{Definitions/mdpi} 

\firstpage{1} 
\pubvolume{1}
\issuenum{1}
\articlenumber{0}
\pubyear{2021}
\copyrightyear{2020}
\datereceived{} 
\dateaccepted{} 
\datepublished{} 
\hreflink{https://doi.org/} 
\pdfoutput=1

\usepackage{color}
\usepackage[T1]{fontenc}
\usepackage{textcomp}
\usepackage{eurosym}
\usepackage{xspace}
\usepackage{soul}
\usepackage{algorithm}
\usepackage{algcompatible}
\usepackage{graphicx}
\usepackage{booktabs}
\usepackage{amsmath,amsfonts,amssymb}

\Title{Flexibility management with virtual batteries of thermostatically controlled loads: real-time control system and potential in Spain}

\TitleCitation{Title}

\Author{Alejandro Mart\'in-Crespo $^{1,}$*\orcidA{}, Sergio Saludes-Rodil $^{1,}$\orcidB{} and Enrique Baeyens $^{2,}$\orcidC{}}

\AuthorNames{Alejandro Mart\'in-Crespo, Sergio Saludes-Rodil and Enrique Baeyens}

\AuthorCitation{Mart\'in-Crespo, A.; Saludes-Rodil, S.; Baeyens, E.}

\address{%
$^{1}$ \quad Centro Tecnol\'ogico CARTIF, Parque Tecnol\'ogico de Boecillo 305, Boecillo, Valladolid, Spain; sersal@cartif.es\\
$^{2}$ \quad Instituto de las Tecnolog\'{\i}as Avanzadas de la Producci\'on, Universidad de Valladolid, Paseo del cauce 59, Valladolid, Spain; enrbae@eii.uva.es}

\corres{Correspondence: alemar@cartif.es; Tel.: +34 983 546 504}

\abstract{Load flexibility management is a promising approach to face the
problem of balancing generation and demand in electrical grids. This problem is
becoming increasingly difficult due to the variability of renewable energies. 
Thermostatically controlled loads can be aggregated and managed by a 
virtual battery, and they provide a cost-effective and efficient alternative
to physical storage systems to mitigate the inherent variability of renewable
energy sources. But virtual batteries require of an accurate control system
being capable of tracking frequency regulation signals with minimal error.  
A real-time control system allowing virtual batteries to accurately track
frequency or power signals is developed. The performance of this controller is
validated for a virtual battery composed of 1,000 thermostatically controlled
loads. Using virtual batteries equipped with the developed controller, a 
study focused on residential thermostatically controlled loads in Spain is 
performed. The results of the study quantify the potential of this technology
in a country with different climate areas and provides insight about the
feasibility of virtual batteries as enablers of electrical systems with 
high levels of penetration of renewable energy sources.}


\begin{document}

\section{Introduction} \label{introduction}

A sustainable and environmentally friendly electric system requires 
increasing the generation of energy coming from renewable sources, such as
solar and wind energy. 
The higher penetration of renewable energy sources in the electric system
reduces emissions of greenhouse gases and pollutants. 
However, this type of energy is uncertain and intermittent, and therefore not
completely predictable. Consequently, the task of balancing electricity demand
and generation is becoming more complicated. As the penetration of
renewable energy sources grows, more advanced and efficient regulation
strategies are required to balance the power of the system. In this scenario,
flexibility management (FM) could become a suitable cost-effective alternative
to overcome these technical difficulties.

Nowadays, an increasing number of houses have electric appliances that
exploit their own thermal inertia, or that of the home, for their operation. 
These appliances are called ``thermostatically controlled loads'' (TCLs), and
they are important FM resources. Their operation can be controlled in order to
provide frequency regulation to the electrical grid,
while respecting their own physical constraints and the user requirements. 
TCLs are also available in many industrial processes.
Some examples are cold rooms and certain chemical reactors. In order to promote
their operation as balancing resources for distribution system operators, TCLs
can be aggregated into virtual batteries (VBs). Likewise conventional
batteries, they are defined in terms of global quantities such as capacity,
state of charge (SOC) and power, but with some differences, as it is explained
later in this paper.
Several studies characterize VBs involving homogeneous or heterogeneous TCLs
\cite{Callaway2011achieving,koch2011modeling,HHao1,acharya2017coordinated,tindemans2015decentralized,wang2020flexibility}.
The last one is closer to the real-world, and has a greater significance and a
broader application possibilities.

The main contribution of this paper is a study about the potential of VBs
enabled by residential TCLs in Spain. Similar studies have been carried out in
other countries and regions, such as  Denmark, \cite{6938939}, Germany,
\cite{GILS2016401}, Great Britain, \cite{trovato2016advanced}, Switzerland
\cite{kamgarpour2014population}, California, \cite{hao2015potentials} and
Sardinia, \cite{conte2017stochastic}. But, to the best of our knowledge, this
is the first one applied to Spain. In all the abovementioned studies, authors
found that TCLs can provide huge potential to mitigate the negative effects of
the increase in renewable energy sources in the electrical system. The current
study for the Spanish case considers three different climate areas
\cite{sech2011analisis}, featuring different prevalent weather conditions that
may affect the renewable energy generation potential and TCLs performance. In
order to fully understand the potential impact of the VBs in Spain, an accurate
real-time VB control system has been developed and validated by simulation. The
control set-point is determined by the grid operator, who determines the power
required at every time instant based on the grid status. 




Flexible loads, such as electrical vehicles (EVs) or TCLs, have been extensively
studied as elements that could be aggregated to act as a virtual battery 
\cite{han2011estimation,liu2013decentralized,kempton2008test}.
A theoretical characterization of the aggregate power and energy capacities for
a collection of TCLs is reported in \cite{HHao1}, as well as a
priority-stack-based control strategy for frequency regulation using TCLs. 
A generalized battery model that can be directly
controlled by an aggregator is introduced in \cite{HHao2}.
The two abovementioned papers establish a TCL behaviour model
and a VB control system. A study that experimentally models VBs is 
reported in \cite{nandanoori2019identification}, where a first-order model is
adjusted using binary search algorithms. The operating reserve provided by an
aggregation of heterogeneous TCLs can also be modelled probabilistically, 
as explained in
\cite{DING201946,lu2005modeling,zhang2012aggregate,perfumo2012load,kundu2011modeling},
using Markov chains, as in \cite{7995082,mathieu2012state,xia2019hierarchical},
or applying model predictive control \cite{liu2015model,liu2019trajectory}.
Several of these papers do not take into account the real short-term
instantaneous status of TCLs, and some of them are focused on changing the
set-point temperature of the devices, which could imply reducing users' comfort.
In \cite{SKhan}, a stochastic battery model along with a control
system which performs TCLs monitoring tasks is proposed, although frequent 
data acquisition is performed, the system does not accurately fit the grid 
requirements. 

In \cite{8586247}, the technical viability of frequency regulation by TCLs is
analysed, and it is proved that the load contribution can be significant.
These thermal devices can also be applied to voltage control.
\cite{bogodorova2016voltage}. TCLs are useful in microgrids, as explained in
\cite{8606271}, and in virtual energy storage systems (VESS)
\cite{cheng2017benefits}, where their capability to provide ancillary 
services is studied.
Another use of TCLs is the management of intraday wholesale energy market
prices \cite{mathieu2014arbitraging}.

The remainder of this paper is organised as follows. 
In Section~\ref{virtualbattery}, models of an individual TCL and a VB 
are introduced. These models are used to develop a new VB control system 
that improves the accuracy in power delivery as compared to previously 
designed control systems.
In Section~\ref{casestudiedspain}, our control systems are used to 
study the capabilities of VBs as a cost-effective solution to balance 
power generation and consumption in electric systems with a large penetration
of renewable energy sources. The study is carried out for residential TCLs
in Spain.
Section~\ref{results} shows and discuss the results of the study. Previously,
the performance of the new VB control system is validated through simulated
experiments.
Finally, the conclusion is given in Section~\ref{conclusion}.

\section{The Virtual Battery} \label{virtualbattery}

A VB is defined as an aggregated collection of TCLs operated by a suitable
control system to provide power and frequency regulation to the electrical
grid. VBs are combined with adequate regulation policies to increase
penetration of renewable energy generation sources in the grid by balancing
power demand and generation in a cost-effective manner.

A new real-time VB control system, which accurately follows the operator
signal, is developed in this section. The control strategy complies with users
and device constraints, such as TCL temperature bounds or short-cycling
prevention. The goal is achieved by anticipating variations imposed by the
abovementioned constraints.  Our control strategy is an improvement of that
reported in \cite{Yo2018}.

The models of TCL and VB used to develop the new control system are explained
below. They quantify the stored energy and the maximum power the VB may
provide. The models are represented in discrete-time and the sampling time 
is denoted by $h$.

\subsection{The TCL model}

A model allows the controller to predict the behavior of each TCL and
estimate its stored energy. The parameters describing a TCL are collected in
Table~\ref{interparameters}. The nameplate power $P_i$ is a positive
quantity in cooling devices and negative in heating devices. The typical values
of these parameters for residential TCLs are shown in Table~\ref{typicalpara}.
Reversible heat pumps are classified into two different types: 'reversible heat
pumps (heat)' and 'reversible heat pumps (cold)'. The motivation for this
classification is that some reversible heat pumps are only used for heating or
cooling, not for both tasks. Besides, they have a different operation mode in
the VB control algorithm.

\begin{table}
\centering
\resizebox{\ifdim\width>\columnwidth\columnwidth\else\width\fi}{!}{
\begin{tabular}{cl}
\toprule
\textbf{Symbol} & \textbf{Meaning} \\ \midrule
$\theta$ & TCL temperature (${^\circ}\mathrm{C}$) \\ 
$\hat\theta_a$ & Forecast ambient temperature (${^\circ}\mathrm{C}$) \\ 
$\theta_s$ & Set point temperature (${^\circ}\mathrm{C}$) \\ 
$\Delta$ & Temperature dead-band (${^\circ}\mathrm{C}$) \\ 
$\omega$ & Perturbation (${^\circ}\mathrm{C}$) \\ 
$R_{th}$ & Thermal resistance (${^\circ}\mathrm{C}$/kW)\\ 
$C_{th}$ & Thermal capacity (kWh/${^\circ}\mathrm{C}$) \\ 
$P$ & Nameplate power (kW) \\ 
$P_0$ & Mean power (kW) \\ 
$\eta$ & Coefficient of performance \\ 
$\phi$ & Kind of device\\ 
$u$ & Status \\ 
$\delta$ & Availability \\ 
$\gamma$ & Full availability \\ 
$\lambda$ & Time to reach bound temperature (h) \\ 
$\zeta$ & Cycle elapsed time (h) \\ 
$\kappa$ & Minimum cycle elapsed time (h) \\ \bottomrule
\end{tabular}
}
\caption{TCL parameters}
\label{interparameters}
\end{table}

\begin{table*}
\centering
\resizebox{\ifdim\width>\textwidth\textwidth\else\width\fi}{!}{
\begin{tabular}{ccccccc}
\toprule
\bf TCL & \boldmath $R_{th}$ (${^\circ}$C/kW) & \boldmath $C_{th}$ (kWh/${^\circ}$C) & \boldmath $P$ (kW) & \boldmath $\eta$ & \boldmath $\theta_s$ (${^\circ}$C) & \boldmath $\Delta$ (${^\circ}$C) \\ \midrule
Reversible heat pump (heat) & $1.5$ -- $2.5$ & $1.5$ -- $2.5$ & $(-4)$ -- $(-7.2)$ & $3.5$ & $15$ -- $24$ & $0.25$ -- $1.0$ \\ 
Reversible heat pump (cold) & $1.5$ -- $2.5$ & $1.5$ -- $2.5$ & $4$ -- $7.2$ & $2.5$ & $18$ -- $27$ & $0.25$ -- $1.0$  \\ 
Non-reversible heat pump & $1.5$ -- $2.5$ & $1.5$ -- $2.5$ & $(-4)$ -- $(-7.2)$ & $3.5$ & $15$ -- $24$ & $0.25$ -- $1.0$ \\ 
Cold pump & $1.5$ -- $2.5$ & $1.5$ -- $2.5$ & $4$ -- $7.2$ & $2.5$ & $18$ -- $27$ & $0.25$ -- $1.0$ \\ 
Electric water heater & $100$ -- $140$ & $0.2$ -- $0.6$ & $(-4)$ -- $(-5)$ & $1$ & $43$ -- $54$ & $2$ -- $4$ \\ 
Refrigerator & $80$ -- $100$ & $0.4$ -- $0.8$ & $0.1$ -- $0.5$ & $2$ & $1.7$ -- $3.3$ & $1$ -- $2$ \\ \bottomrule
\end{tabular}
}
\caption{Range of values for the parameters of residential TCLs~\cite{osti_1182734}}
\label{typicalpara}
\end{table*}

Several discrete-time models of a TCL have been proposed in the literature
\cite{HHao1,SKhan}. Here, we use the model given in \cite{SKhan} because
it is more realistic with a small cost of increasing complexity.
In this model, the temperature $\theta_i$ of each TCL $i$ is calculated 
for every time instant $k$ using the following equation
\begin{equation}\label{eq:mod1}
\theta_{i}^{k+1} = 
g_{i} \theta_{i}^{k} + 
(1 - g_{i})(\hat\theta_{a_{i}}^{k} - 
u_{i}^{k} \theta_{g_{i}}) + \omega_{i}^{k},
\end{equation}
where
\begin{equation}\label{eq:mod2}
g_{i} = e^{-1/{R_{th_{i}} C_{th_{i}}}}, \quad
\theta_{g_{i}} = R_{th_{i}} P_{i} \eta_{i}.
\end{equation}

Each TCL has a set-point temperature ($\theta_{s_i}$), which is set by 
the user, and a temperature dead-band ($\Delta_i$) determined by the design 
of the TCL or by the user. 
Both parameters define the comfort band, whose bounds are
$\theta_{s_i} \pm \Delta_i$. The temperature $\theta_i$ must always be
inside this comfort band.

Consequently, the control system aims to maintain the temperature $\theta_i$ 
in the comfort band.
The average power to get the objective ($P_{0_i}$) is given by
\begin{equation}
P_{0_{i}}^{k} = \frac{\hat\theta_{a_{i}}^{k}-\theta_{s_{i}}}{\eta_i R_{th_i}}.
\end{equation}

In order to maintain the temperature in the comfort band, each load can be 
switched on and off by the control system depending on the temperature and its 
own status.

Each device is characterized by four binary-valued variables:
device type ($\phi_i$), status ($u_i$), availability ($\delta_i$) and total
availability ($\gamma_i$). 
The device type $\phi_i$ is 0 for a cooling TCL and 1 for a heating TCL. 
The device status $u_i$ is 1 when it is working, \emph{i.e.}, the
motor is on with fixed power consumption $P_i$, and 0 when off,
with zero power consumption. 
The device availability $\delta_i$ is 1 when the TCL is available for 
the control system and 0 otherwise.
A TCL is available when $\theta_i$ lies within the comfort band, and the
time elapsed after its last change of status is large enough to avoid 
short-cycling.
The full availability of the device $\gamma_i$ is 0 when the ambient
forecast temperature $\hat\theta_{a_i}$ is lower (in cooling devices) 
or higher (in heating devices) than any TCL temperature bound.
This last parameter indicates whether the TCL is ready to operate or not.

Short-cycling has harmful effects for electrical TCL systems. It causes
damage in the electromechanical components, decreases their expected lifetime 
and can also significantly drive up the energy consumption.
In order to avoid short-cycling, a minimum cycle elapsed time $\kappa_i$ 
is defined. Let $\zeta_i$ be the elapsed time after the last status change, 
then the TCL device is available if 
$\theta_i \in [\theta_{s_i}-\Delta_i,\theta_{s_i}+\Delta_i]$
and $\zeta_i > \kappa_i$. 
 
The time to reach the temperature bound ($\lambda_i$) is the time that a TCL 
can be off without reaching that bound.
It is calculated by simulating the evolution of $\theta_i$ over time using the
TCL dynamic model given by \eqref{eq:mod1}. 
If the TCL is not available,
\emph{i.e.} if $\delta_i = 0$, then $\lambda_i$ also equals $0$.

\subsection{The VB model} \label{sectionVB}

The VB model is an abstraction that represents a large number of TCLs with
a reduced number of parameters.  It allows the grid operator to make 
decisions in order to efficiently manage the grid. The set of parameters that
describe a VB is given in Table~\ref{vbparameters}.

\begin{table}
\centering
\resizebox{\ifdim\width>\columnwidth\columnwidth\else\width\fi}{!}{
\begin{tabular}{cl}
\toprule
\textbf{Symbol} & \textbf{Meaning} \\ \midrule
$\mathcal N$ & Set of TCLs with cardinality $N$\\ 
$C_c$/$C_d$ & Charging/Discharging capacity (kWh) \\ 
$SOC_c$/$SOC_d$ & Charging/Discharging state of charge (kWh) \\ 
$n_+$/$n_-$ & Maximum charging/discharging power (kW) \\ 
$n'_+$/$n'_-$ & Maximum available \\
              & charging/discharging power (kW) \\ \bottomrule
\end{tabular}
}
\caption{VB parameters}
\label{vbparameters}
\end{table}

The capacity of the VB is represented by two different parameters, 
$C_c$ and $C_d$, to model the lack of symmetry in the dynamics of 
the charging and discharging processes of a TCL.
In one case, the temperature change is forced by mechanical or electrical
components (increasing $\theta$ in heating devices or reducing it in cooling
ones), and in the other case it is not. Likewise, the state of charge 
($SOC_c$ and $SOC_d$), the maximum power ($n_+$ and $n_-$), and the maximum 
available power ($n'_+$ and $n'_-$) are also duplicated.

The charging capacity ($C_c$) is the maximum energy that can be used from every
TCL $i\in\mathcal N$ whose status $u_i$ is 1. 
If the VB is charging, then the charging state of charge ($SOC_c$) is the 
current energy reserve of the VB for charging. 
Likewise, the discharging capacity ($C_d$) and the discharging state of charge
($SOC_d$) represent the maximum energy that can be used for discharging and the
current energy reserve for discharging.
The capacities $C_c$ and $C_d$ are calculated by using the models to
obtain the time that total available TCLs 
(\emph{i.e.}, $\{i\in\mathcal N:\gamma_i=1\}$) take to evolve from one 
temperature bound to the other. The states of charge
$SOC_c$ and $SOC_d$ also use the TLC models, but operating from the
actual temperature $\theta_i$ to the corresponding temperature bound. 
The algorithms developed for computing the capacities and states of charge
require long-term temperature forecasts $\hat\theta_{a_i}$ to produce
accurate results. Accurate forecasts are assumed to be available in this paper.

The maximum charging power ($n_+$) is the maximum power a VB can provide for
charging, assuming that all TCLs are available ($\delta_i$ is 1):
\begin{equation}
n_{+}^{k+1} = \sum_{i=1}^{N} (1-2\phi_{i})(P_{i}-P_{0_{i}}^{k}).
\end{equation}

The maximum available charging power ($n'_+$) is the maximum power that a VB
can provide in the following instant, considering only the available TCLs.
It is calculated by excluding the TCLs that are unavailable for
charging:
\begin{equation}
(n'_{+})^{k+1} = n_{+}^{k+1} - P_{+}^{k},
\end{equation}
where $P_{+}$ is the sum of power of the TCLs which are unavailable for 
charging. TCLs are unavailable for charging when their status cannot be 
on in the next time instant.

The maximum available discharging power ($n_-$) is the same as $n_+$, but for 
the discharging process:
\begin{equation}
n_{-}^{k+1} = \sum_{i=1}^{N} (1-2\phi_{i})(P_{0_{i}}^{k}).
\end{equation}

The maximum available discharging power ($n'_-$) excludes the power of the
TCLs which are unavailable
\begin{equation}
(n'_{-})^{k+1} = n_{-}^{k+1} - P_{-}^{k},
\end{equation}
where $P_{-}$ is the sum of power of TCLs that are unavailable for discharging.
TCLs are unavailable for discharging when their status cannot be off in the
following instant.

The maximum powers $n_+$, $n_-$, $n'_+$ and $n'_-$ are calculated considering 
only TCLs that are totally available ($\gamma_i$ is 1).

\subsection{The VB controller}

\begin{figure}
\centering
\resizebox{\ifdim\width>\columnwidth\columnwidth\else\width\fi}{!}{
\includegraphics[width=0.50\textwidth]{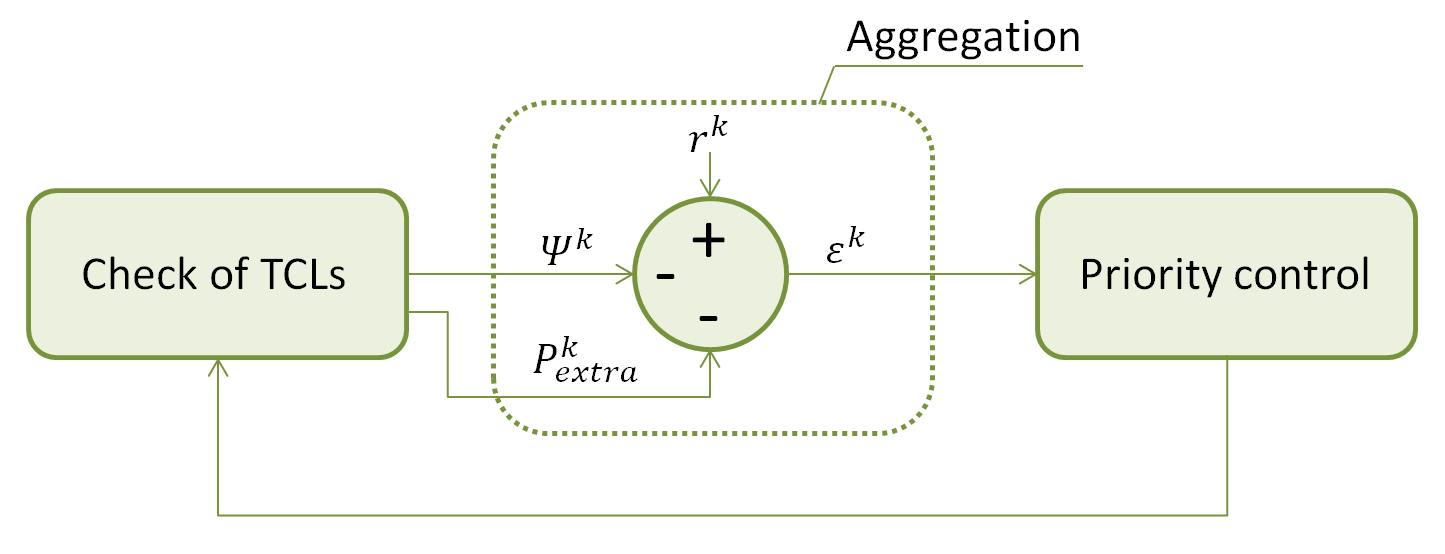}
}
\caption{Structure of the VB controller}
\label{VBcontroller}
\end{figure}

The real-time VB control algorithm is composed of three blocks:
\textit{Check of TCLs}, \textit{Aggregation}, and \textit{Priority Control}. 
The first block is executed individually by each TCL, while the others are
managed by a centralised controller. The communication between TCLs and the
aggregator is done by using the method explained in \cite{LAKSHMANAN2016705},
which requires an internet connection. An accurate long-term weather forecast 
is required to obtain reliable results.  
Figure~\ref{VBcontroller} shows the complete controller structure where the
interconnection of the three blocks is displayed. The set of variables
that are used in the control algorithm are listed in Table~\ref{vbvariables}.

\begin{table}
\centering
\resizebox{\ifdim\width>\columnwidth\columnwidth\else\width\fi}{!}{
\begin{tabular}{cl}
\toprule
\textbf{Symbol} & \textbf{Meaning} \\
\midrule
$P_{extra}$ & Extra power (kW) \\
$r$ & System operator signal (kW) \\
$P_{agg}$ & Aggregated power (kW) \\
$P_{base}$ & Base power (kW) \\
$\psi$ & System deviation (kW) \\
$\epsilon$ & Regulation signal (kW) \\
$p$ & Power switched (kW) \\
\bottomrule
\end{tabular}
}
\caption{VB controller variables}
\label{vbvariables}
\end{table}

\subsubsection{Checking of TCL}

The objective of this block is to compute the variables of a TCL
at the next time instant and to avoid its temperature from leaving
the comfort band. The process that a TCL $i$ executes is detailed in
Algorithm~\ref{algoritmo1}, see the Appendix. 
In this algorithm, a TCL with variables that are close to violating their 
constraints are identified and marked, which means that $\delta_i$ and/or
$\gamma_i$ become 1. The powers $P_{+}$, $P_{-}$ and the extra power 
($P_{extra}$) are calculated as well. 
$P_{extra}$ is the VB total power that will be switched on or off at the
next time instant in order to avoid violating the TCLs constraints.

The computed variables of each TCL should be automatically communicated to
the centralised controller every time instant $h$.
If the length of the sampling time interval $h$ is too small, then
the communication system would require a high bandwidth and very
expensive hardware due to the high volume of data. Notwithstanding,
the variables must be sent as often as possible, so the controller may
quickly correct any power deviation. In our studies, we have
used $h=10$ seconds.

\subsubsection{Aggregation}

In this controller block, the data obtained from each device are 
aggregated and compared to the system operator signal ($r$). 
This value is the amount of power that the grid operator wants to be provided.
It can be either positive or negative, and has been previously determined by
the grid operator using the current grid status and information about
the VB capacity, state of charge and maximum power. If $r$ is
positive, then the VB is ordered to absorb energy. It could be caused by an
excess in the renewable energy production or a strong power demand fall. 
However, if $r$ is negative, then the VB has to stop consuming energy. 
This case corresponds to a higher power demand or a lower electric generation
than expected.

System deviation ($\psi$) is the difference between the current power consumed
by the VB, and its base power. The system deviation is obtained as
\begin{equation}
\psi^{k}=P_{agg}^{k}-P_{base}^{k},
\end{equation}
where
\begin{equation}
P_{agg}^{k}=\sum_{i=1}^{N} (1 - 2\phi_{i}) u_{i}^{k} P_{i},
\end{equation}
\begin{equation}
P_{base}^{k}=\sum_{i=1}^{N} (1 - 2\phi_{i}) P_{0_{i}}^{k}.
\end{equation}

As a result of the aggregation process, the regulation signal ($\epsilon$) 
is given by
\begin{equation}
\epsilon^{k} = r^{k} - \psi^{k} - P_{extra}^{k}.
\end{equation}

This variable will be used at the Priority Control block, and it determines 
the power that finally has to be provided.
The aggregation block is implemented in Algorithm~\ref{algoritmo2}. 
See the Appendix.

\subsubsection{Priority Control}

The last block of the control system decides which TCLs change their status
variables $u_i$ in such a way that the output VB power tracks the 
system operator signal $r$. 
The selected TCLs must be available and its temperature be far enough from 
the temperature bounds. The decision making process has been implemented in
Algorithm~\ref{algoritmo3}, see the Appendix.
The variable $p$ tracks the total power obtained during the execution of 
the algorithm. 
The process stops when the absolute value of $p$ exceeds the absolute value of
$\epsilon$ or when every TCL has been assessed. When a TCL changes its status
$u_i$, then the cycle elapsed time $\zeta_i$ is reset. The algorithm only works
properly when the requirements of the grid operator are between the previously
calculated capacities and the maximum available power.
In that case, the maximum error between $\psi$ and $r$ is the greatest $P_i$ 
in the aggregation of TCLs.

\section{Residential Virtual Battery Potential in Spain}
\label{casestudiedspain}

\begin{figure}
\centering
\resizebox{\ifdim\width>\columnwidth\columnwidth\else\width\fi}{!}{
\includegraphics[width=0.5\textwidth]{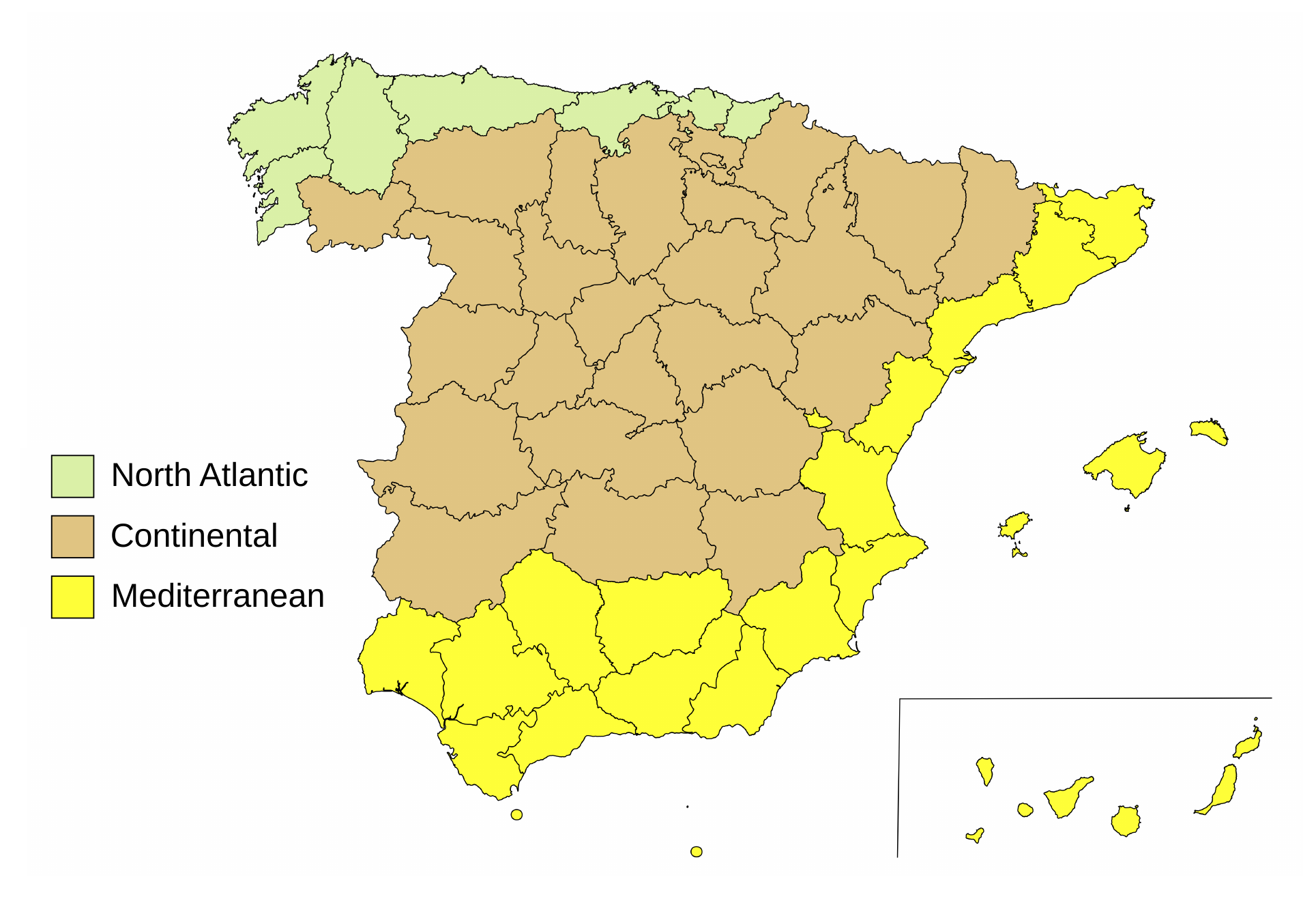}
}
\caption{Territorial distribution of climatic zones in Spain \cite{sech2011analisis}}
\label{MapaEspana}
\end{figure}

The model of a VB and its control system developed in
Section~\ref{virtualbattery} are used here to perform studies about the
implementation of residential VBs and their capabilities.
The study focuses on Spain for several reasons:
its great potential for renewable energy penetration in the electric grid, 
the presence of different climate areas, and the availability of data. 
The main source of information about residential TCLs in Spain 
is the SECH-SPAHOUSEC pro\-ject~\cite{sech2011analisis}. It was
developed by IDAE (Instituto para la Diversificaci\'on y Ahorro de la
Energ\'ia), and it estimates the penetration per house of heat pumps, 
cold pumps, electric water heaters and refrigerators in the different 
climate areas of the country in 2010. 
A summary of that information is shown in
Table~\ref{TCLpenetrationpercent}. According to \cite{sech2011analisis}, Spain
is geographically classified in three different climate areas: North Atlantic,
Continental and Mediterranean, see Figure~\ref{MapaEspana}.
In order to update the current number of TCLs in Spain, shown in 
Table~\ref{TCLnumber2017}, the variation in the number of houses between 
2010 and 2019 in each climate area is considered.
This information is obtained from the ECH survey 
(Encuesta Continua de Hogares) \cite{INEENI}, developed by the Spanish INE (Instituto Nacional de Estad\'istica).

\begin{table*}
\centering
\resizebox{\ifdim\width>\textwidth\textwidth\else\width\fi}{!}{
\begin{tabular}{cccc}
\toprule
\textbf{TCL} & \textbf{North Atlantic} & \textbf{Continental} & \textbf{Mediterranean} \\ 
\midrule
Reversible heat pump (heat) (\%)  & 0.0 & 7.9 & 30.5 \\
Reversible heat pump (cold) (\%)  & 0.3 & 25.9 & 55.4 \\
Non-reversible heat pump (\%)  & 1.5 & 0.9 & 0.7 \\
Cold pump (\%)  & 0.1 & 9.8 & 8.0 \\
Electric water heater (\%)  & 19.9 & 18.1 & 38.0 \\
Refrigerator (\%)  & 99.9 & 99.8 & 99.4 \\
\bottomrule
\end{tabular}
}
\caption{TCLs penetration (\%) in Spain in 2010 \cite{sech2011analisis}}
\label{TCLpenetrationpercent}
\end{table*}

\begin{table*}
\centering
\resizebox{\ifdim\width>\textwidth\textwidth\else\width\fi}{!}{
\begin{tabular}{cccc}
\toprule
\textbf{TCL} & \textbf{North Atlantic} & \textbf{Continental} & \textbf{Mediterranean} \\ 
\midrule
Reversible heat pump (heat) (\#) & 0 & 488924 & 3035739 \\
Reversible heat pump (cold) (\#) & 7444 & 1611424 & 5508641 \\
Non-reversible heat pump (\#) & 36527 & 56542 & 73960 \\
Cold pump (\#) & 1329 & 610388 & 796430 \\
Electric water heater (\#) & 480534 & 1123051 & 3783823 \\
Refrigerator (\#) & 2414583 & 6200175 & 9890698 \\
\bottomrule
\end{tabular}
}
\caption{Number (\#) of TCLs per climate area in 2019}
\label{TCLnumber2017}
\end{table*}

In the North Atlantic climate area, the deployment of cold pumps and reversible
heat pumps for air conditioning is very reduced. The reason is that the 
temperature is not so hot in summer as in other climate areas.

Unfortunately, to the best of our knowledge, there does not exist publicly
available databases concerning thermal and electrical characteristics of TCLs
in Spain. Thus, the study has been performed by sampling random data from
Table~\ref{typicalpara}.

\section{Results and discussion} \label{results}


In this section, two case studies are performed and the obtained results 
are discussed. In the first case study, the VB operation of the controller
is analysed. This study demonstrates the accurate behaviour of the controller.
Later, in a second case study, the residential VB charging capacity,
discharging capacity, maximum charging power, and maximum discharging power of
each Spain climate area are obtained along a natural year, concluding maximum
power ratios and TCL contribution percentages. For this study, a VB controlled
by the new developed controller has been used.

\subsection{Case study: VB controller operation} \label{VBco}

\begin{figure}
  \centering
  \resizebox{\ifdim\width>\columnwidth\columnwidth\else\width\fi}{!}{
    \includegraphics[width=.4\textwidth]{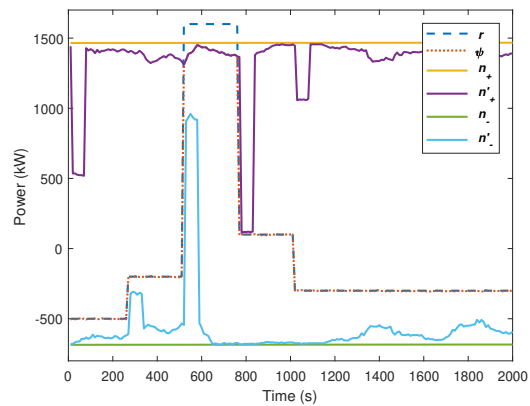}
  }
  \caption{System operator signal and system deviation}
  \label{3VBCO}
\end{figure}

\begin{figure}
  \centering
  \resizebox{\ifdim\width>\columnwidth\columnwidth\else\width\fi}{!}{
    \includegraphics[width=.4\textwidth]{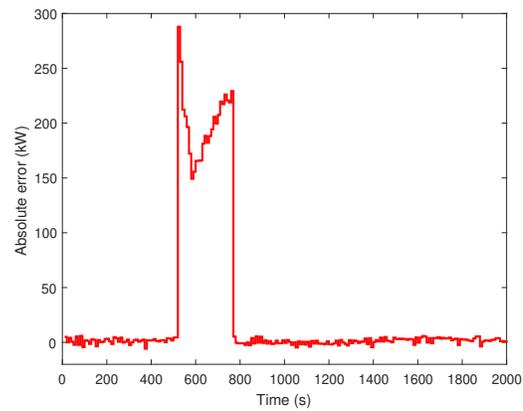}
  }
  \caption{Absolute error following operator signal}
  \label{4VBCO}
\end{figure}

\begin{figure}
  \centering
  \resizebox{\ifdim\width>\columnwidth\columnwidth\else\width\fi}{!}{
    \includegraphics[width=.4\textwidth]{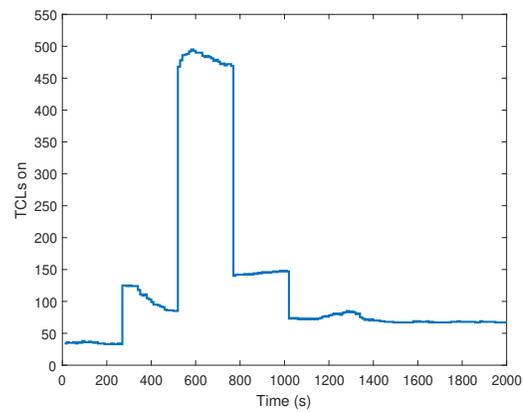}
  }
  \caption{Number of TCLs on}
  \label{5VBCO}
\end{figure}

\begin{figure}
  \centering
  \resizebox{\ifdim\width>\columnwidth\columnwidth\else\width\fi}{!}{
    \includegraphics[width=.4\textwidth]{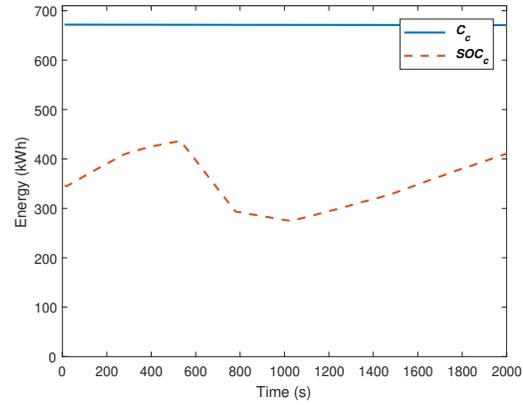}
  }
  \caption{Charging capacity and charging state of charge}
  \label{1VBCO}
\end{figure}

\begin{figure}
  \centering
  \resizebox{\ifdim\width>\columnwidth\columnwidth\else\width\fi}{!}{
    \includegraphics[width=.4\textwidth]{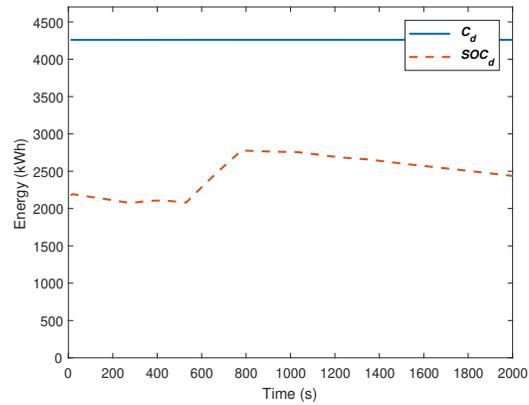}
  }
  \caption{Discharging capacity and discharging state of charge}
  \label{2VBCO}
\end{figure}

\begin{figure}
  \centering
  \resizebox{\ifdim\width>\columnwidth\columnwidth\else\width\fi}{!}{
    \includegraphics[width=.4\textwidth]{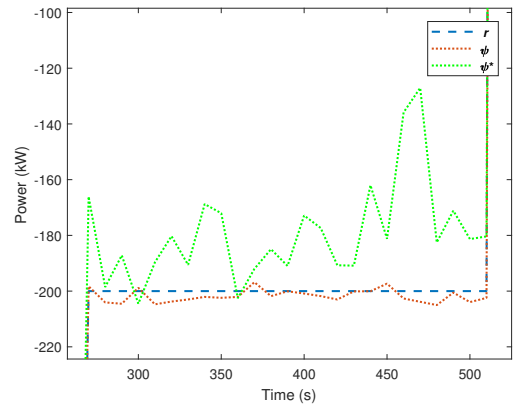}
  }
  \caption{System deviation comparison during instants between seconds 270-510}
  \label{DeviationComparison}
\end{figure}

The control system developed in Section~\ref{virtualbattery}
has been implemented in MATLAB~\cite{MATLAB:2019b} and its performance
is studied for a VB of 1000 TCLs classified as follows: 125 reversible
heat pumps (cold), 125 cold pumps, 125 reversible heat pumps (heat), 125
non-reversible heat pumps, 250 refrigerators and 250 electric water heaters.
The parameters of each TCL have been randomly selected by using a Gaussian 
probability distribution centered on the mean values given in
Table~\ref{typicalpara} and standard deviation $0.1$.
The simulation covers 200 time instants with a sampling interval $h$ of 10 seconds, which amounts to a total of 2000 seconds.
The minimum cycle elapsed time, $\kappa$, is given a value of 60 seconds. No
disturbance has been considered, \emph{i.e.} $\omega=0$. The initial status of
each TCL has also been conveniently randomized using a binary probability
distribution. The estimated ambient temperature $\hat\theta_a$ is fixed at
20${^\circ}$C in case of refrigerators and electric water heaters.  For the
remaining TCLs, $\hat\theta_a$ is variable.
Figures~\ref{3VBCO}--\ref{DeviationComparison} show the obtained results.

\begin{table*}
\centering
\resizebox{\ifdim\width>\textwidth\textwidth\else\width\fi}{!}{
\begin{tabular}{cccc}
\toprule
\bf Greatest capacities and maximum powers & \bf Value \\
\midrule
Greatest charging capacity (kWh) & 671.9\\
Greatest discharging capacity (kWh) & 4260.7\\
Greatest maximum charging power (kW) & 1467.2\\
Greatest maximum discharging power (kW) & 685.7\\
\bottomrule
\end{tabular}
}
\caption{Greatest capacities and maximum powers}
\label{maxcappow}
\end{table*}

The system deviation $\psi$ accurately tracks the system operator
power signal $r$ most of the time. The exception is the time interval 
between 510 and 750 seconds, where the maximum $\psi$ stays at the value 
given by $n'_+$, the maximum available charging power. 
The reason is that the grid operator established a value of $r$ that the VB
is not capable of producing as it is outside the range previously defined by 
$n'_+$ and $n'_-$. This situation should be avoided by taking into account the
capability, SOC and power information of the VB. When the signal $r$ changes,
several TCLs change their status, which means that they become unavailable
for 1 minute. This explains the peaks and valleys in $n'_+$ and $n'_-$ when
$r$ is modified. The absolute error is lower than 6~kW in absolute value
whenever $r$ is between $n'_+$ and $n'_-$. Otherwise, the VB cannot follow the
system operator signal and the absolute error is either $|r-n'_+|$ or
$|n'_--r|$.
The charging/discharging capacities $C_c$/$C_d$ and the maximum 
charging/discharging powers $n_+$/$n_-$ appear to be constant in
Figures~\ref{3VBCO}, \ref{1VBCO} and \ref{2VBCO}. 
In fact, they evolve with time but the changes are very small because the 
simulation time is not large enough to notice great changes in the ambient 
temperatures.
The greatest values of these variables are given in 
Table~\ref{maxcappow}. 
The charging capacity $C_d$ is about 6.5 times greater than the discharging
capacity $C_c$ at every time instant. The reason is the lack of symmetry in
the charging and discharging process, as the TCL takes more time to change
its temperature $\theta$ when it is not forced to by electromechanical 
components, \emph{i.e.}, when it is switched off ($u_i=0$).
Regarding the maximum charging/discharging power, $n_+$ is about twice as
large as $n_-$, since the sum of the nameplate powers $P_i$ doubles the sum of
the average powers ${P_0}_i$ for every TCL $i\in\mathcal N$.

Figure~\ref{DeviationComparison} demonstrates the improvement achieved by the
new controller proposed in this paper. In this figure, $\psi$ is compared to
$\psi^*$ between the seconds 270 and 510, which is the system deviation signal 
when the anticipation to the variations imposed by TCLs constraints is not 
included in the controller. As mentioned above,
$\psi$ follows $r$ with an absolute error not greater than an
individual TCL nameplate power; while $\psi^*$ is several times
greater, as it depends on how many TCLs have to change their status forced by
any of their inner constraints, which is almost random. Specifically, in
this case, the maximum absolute error $\psi^*$ becomes around 15 times as 
higher as the maximum absolute error $\psi$, in absolute value. 

Besides performing short-duration regulation, the VB is able to maintain
a system deviation of -300 kW for more than 15 minutes, as can be seen
from time 1010 to 2000 seconds. This means that this VB can
provide ancillary services to the grid similar to the primary and secondary
regulation, according to the current definition given by the Spanish normative
\cite{BOEref}, which is followed by the grid operator of the country, Red
El\'ectrica de Espa\~na. Using European Network of Transmission System
Operators (ENTSO-E) terminology \cite{SEDCp}, this VB could supply regulation 
similar to Frequency Containment Reserves (FCR) and Frequency Restoration 
Reserves (FRR) services.  
The main difference between the regulation provided by the VB and the 
primary and secondary regulation is that the first acts on the demand side,
whereas the others act on the generation side. 
As a general rule, the VB will fail to follow $r$ earlier if the amount of
power requested is larger.

\subsection{Case study: VB potential in Spain}

\begin{figure}
  \centering
  \resizebox{\ifdim\width>\columnwidth\columnwidth\else\width\fi}{!}{
    \includegraphics[width=.25\textwidth]{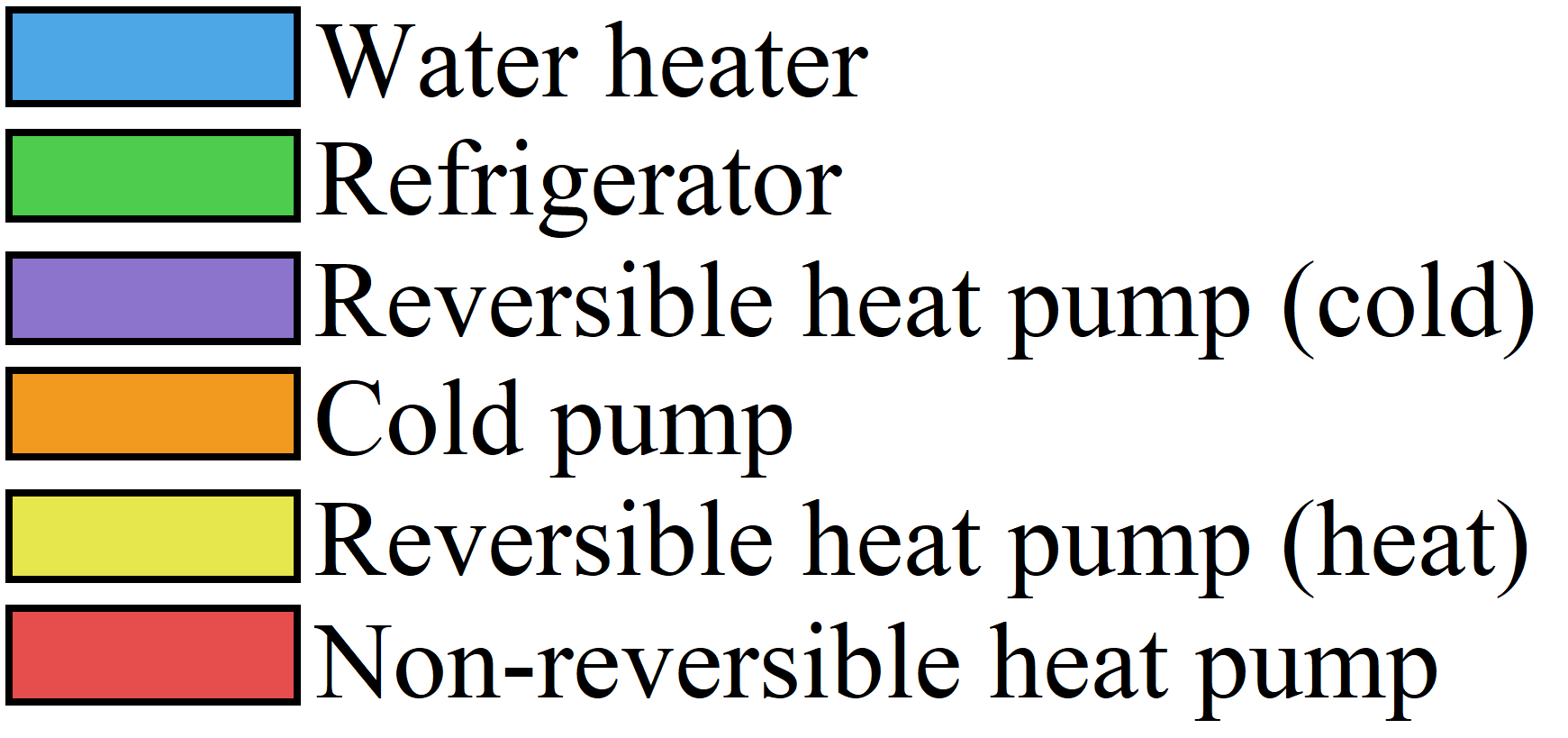}
  }
  \caption{Legend and colour codes for Figures~\ref{fig:regs1}--\ref{fig:regs4}}
  \label{fig:legend}
\end{figure}

\begin{figure}
  \centering
  \resizebox{\ifdim\width>\columnwidth\columnwidth\else\width\fi}{!}{
  \begin{tabular}{c}
    \includegraphics[width=.4\textwidth]{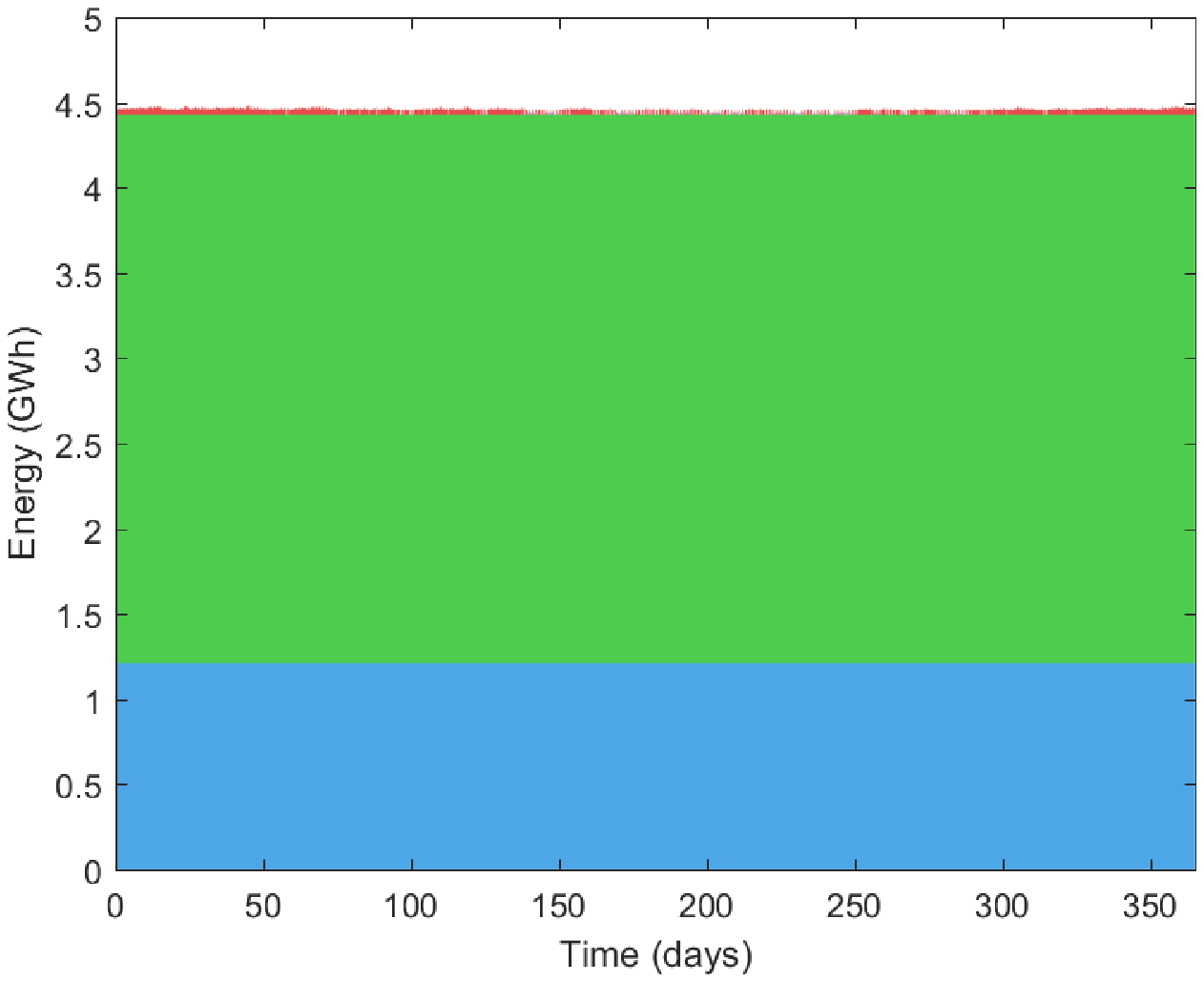}\\
    \includegraphics[width=.4\textwidth]{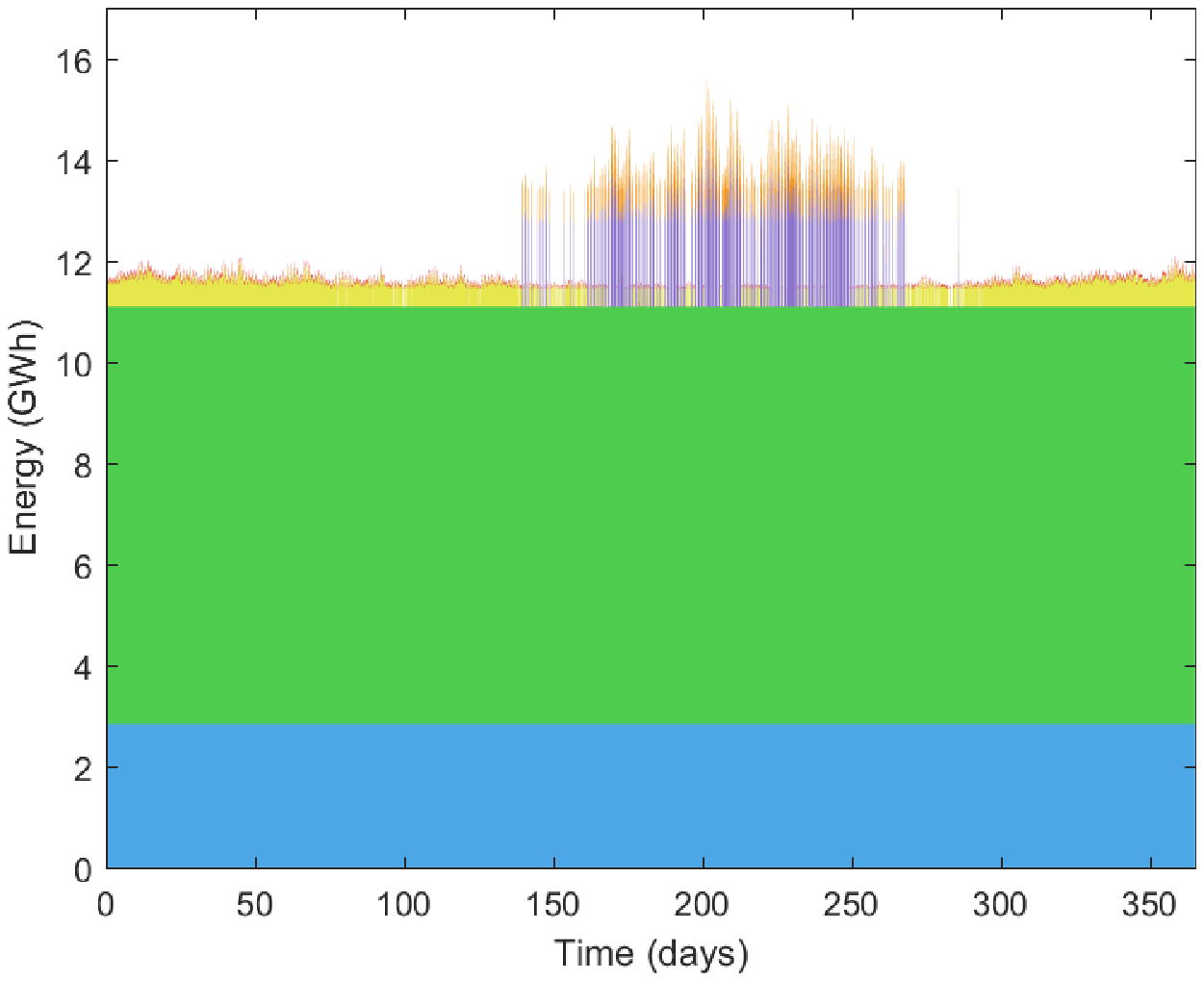}\\
    \includegraphics[width=.4\textwidth]{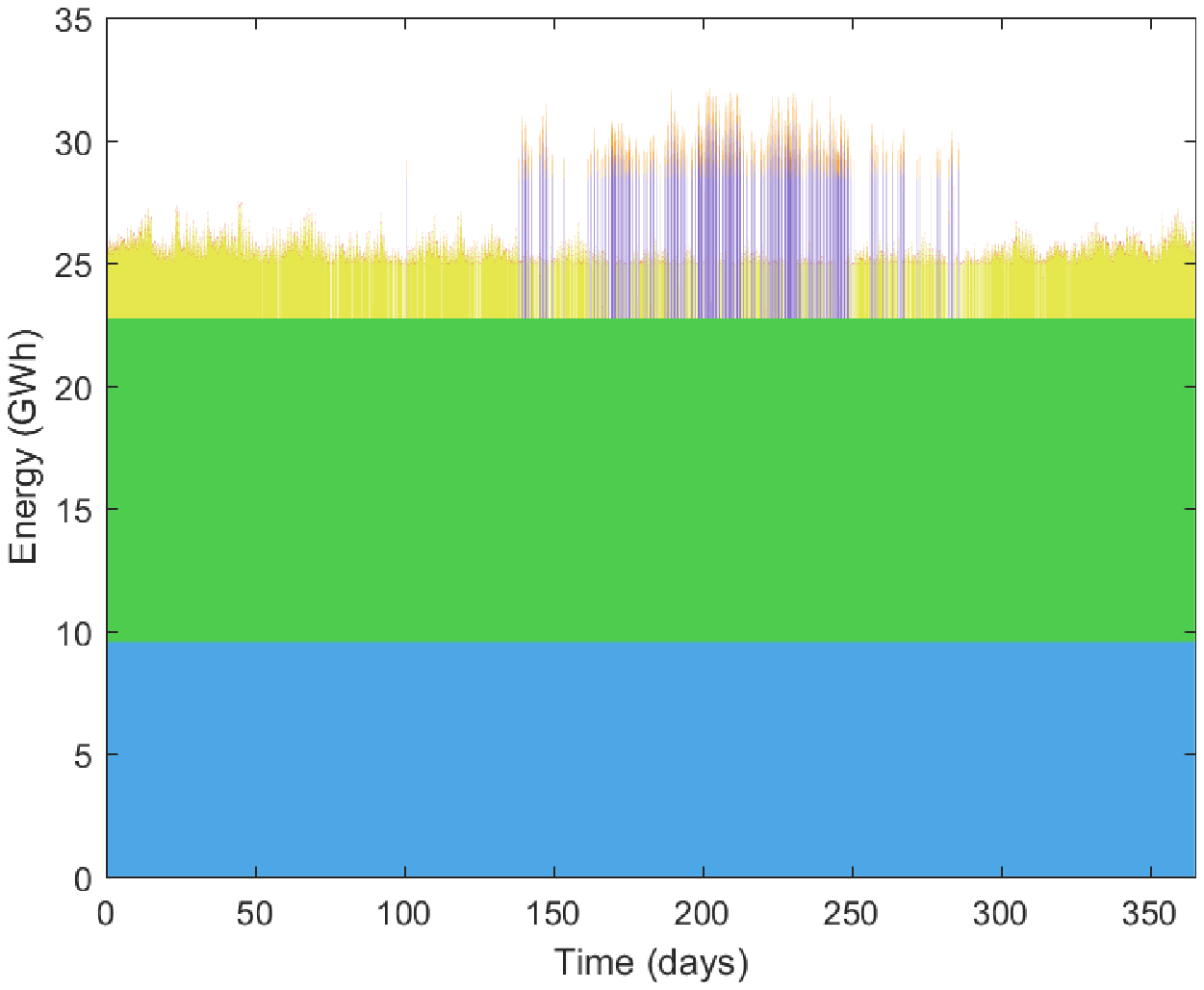}\\
   \end{tabular}
   }
\caption{Charge capacity in North Atlantic, Continental and Mediterranean climate areas in 2019}
\label{fig:regs1}
\end{figure}

\begin{figure}
  \centering
  \resizebox{\ifdim\width>\columnwidth\columnwidth\else\width\fi}{!}{
  \begin{tabular}{c}
    \includegraphics[width=.4\textwidth]{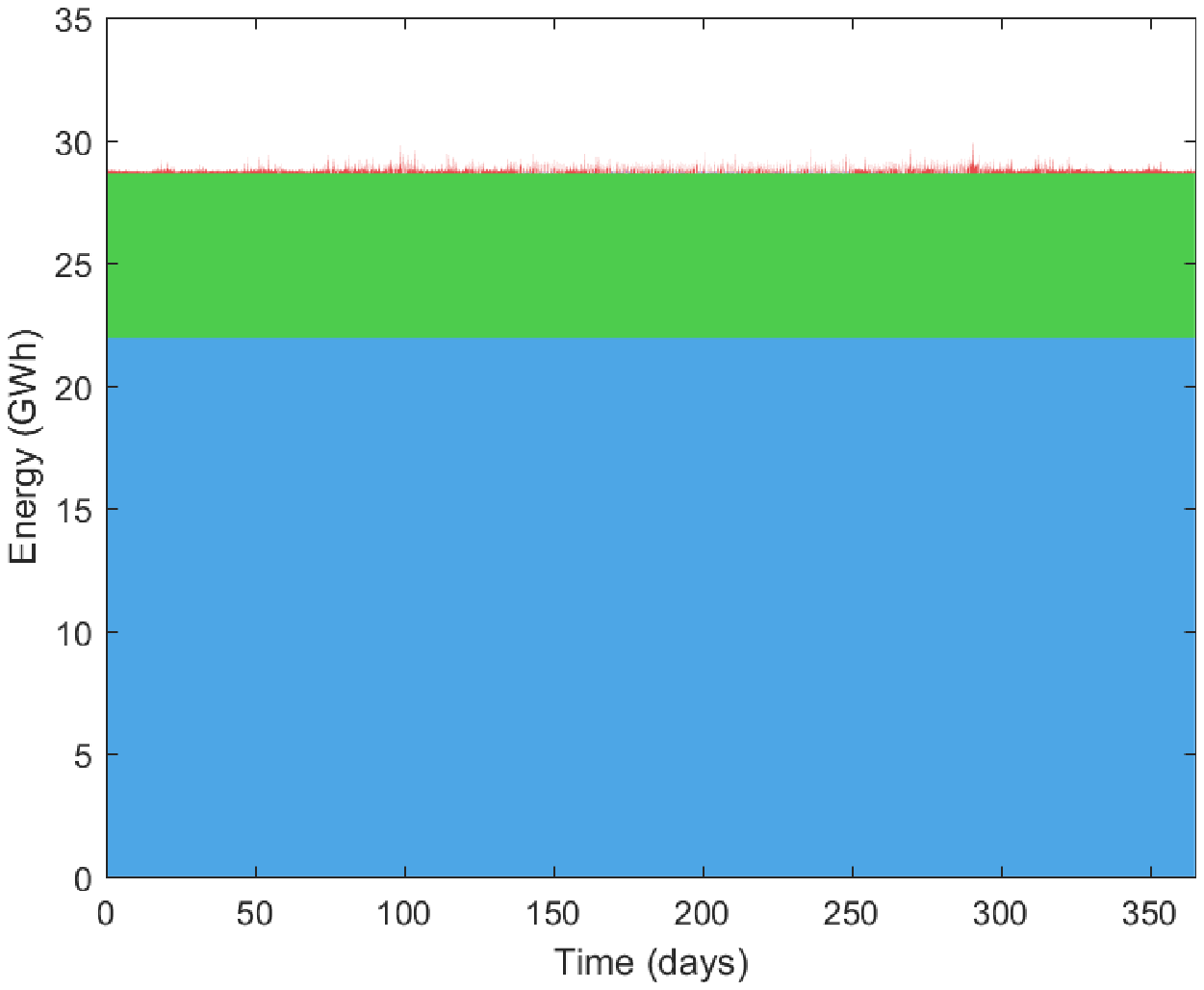}\\
    \includegraphics[width=.4\textwidth]{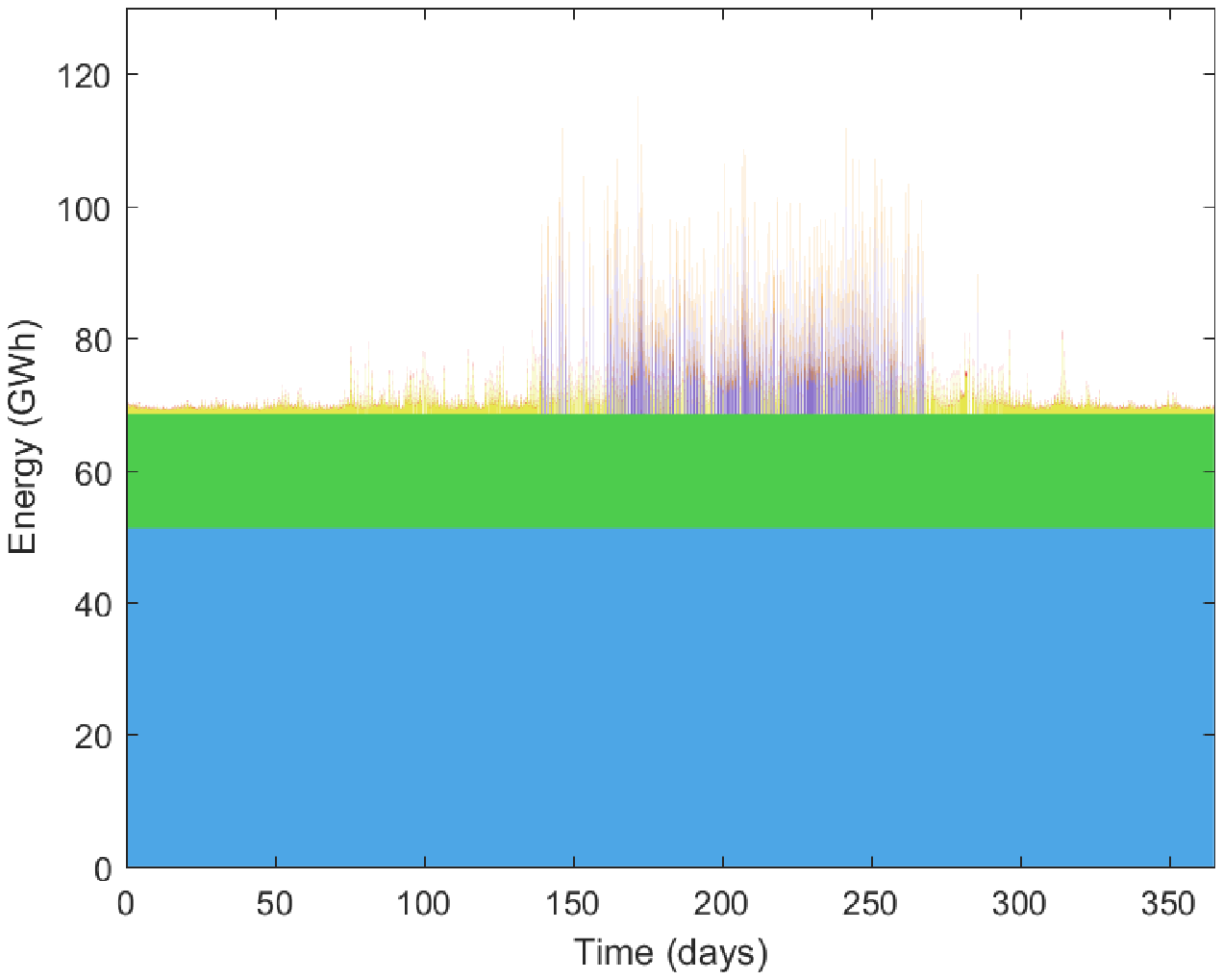}\\
    \includegraphics[width=.4\textwidth]{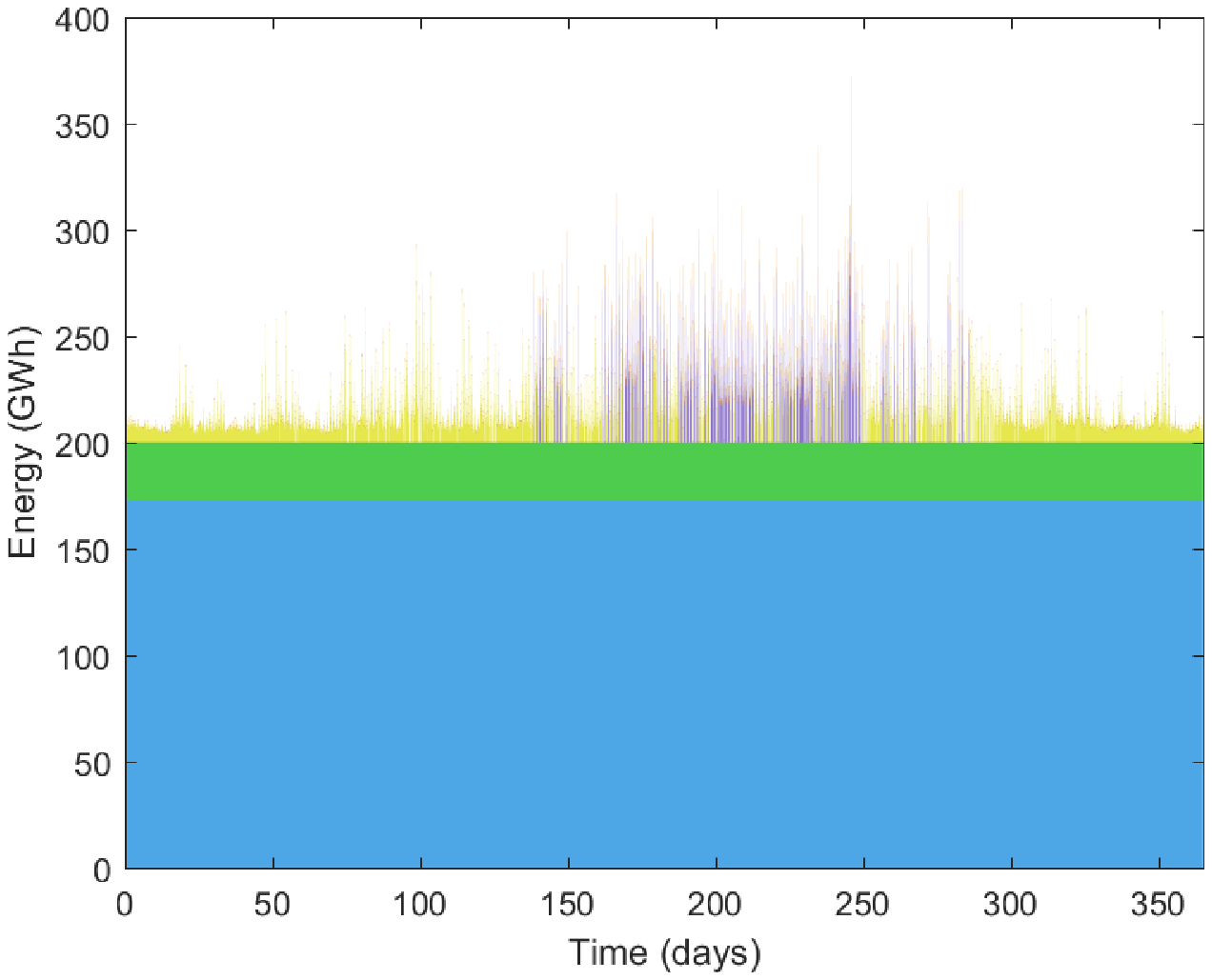}
  \end{tabular}
  }
\caption{Discharging capacity in North Atlantic, Continental and Mediterranean climate areas in 2019}
\label{fig:regs2}
\end{figure}

\begin{figure}
  \centering
  \resizebox{\ifdim\width>\columnwidth\columnwidth\else\width\fi}{!}{
  \begin{tabular}{c}
    \includegraphics[width=.4\textwidth]{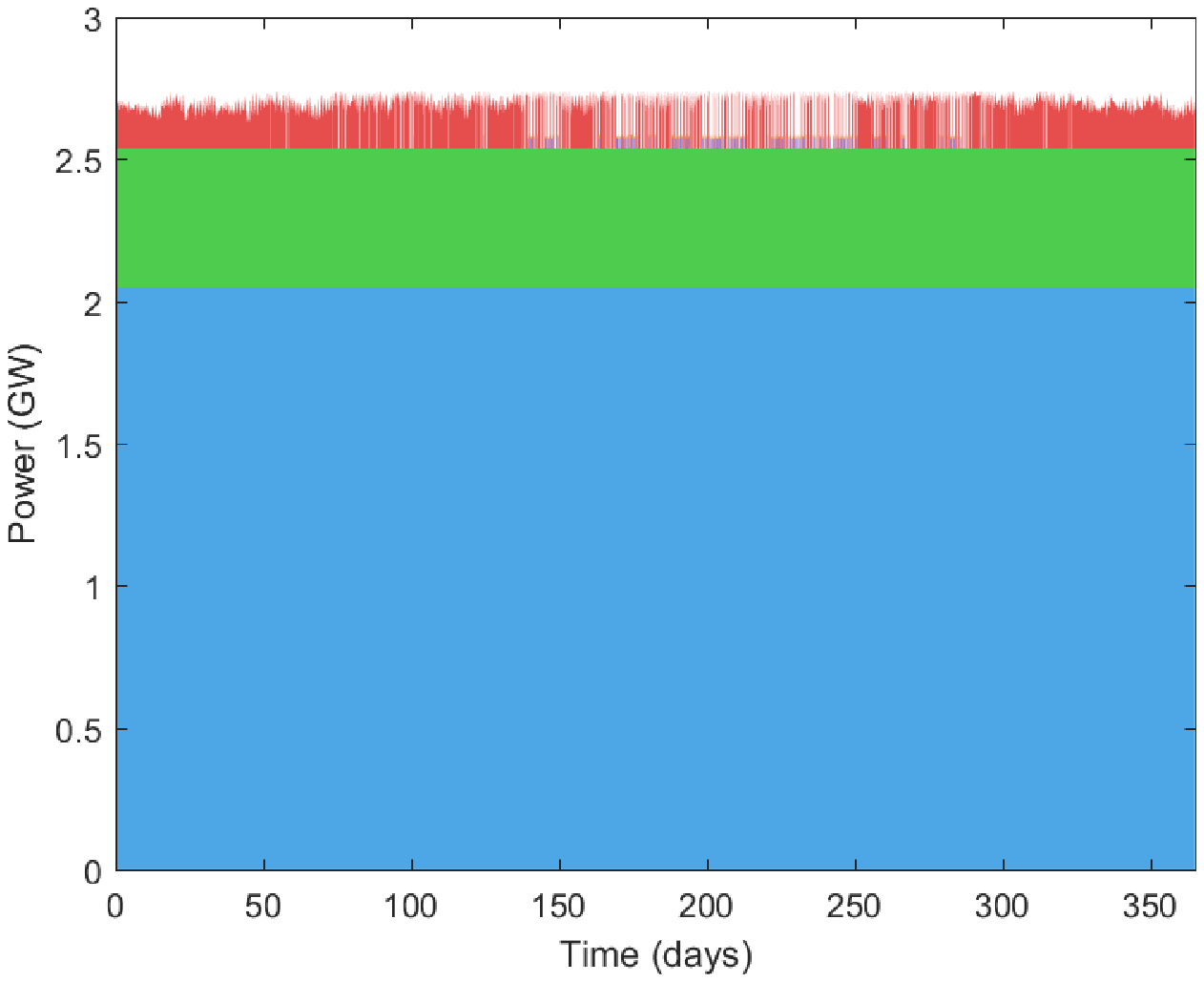}\\
    \includegraphics[width=.4\textwidth]{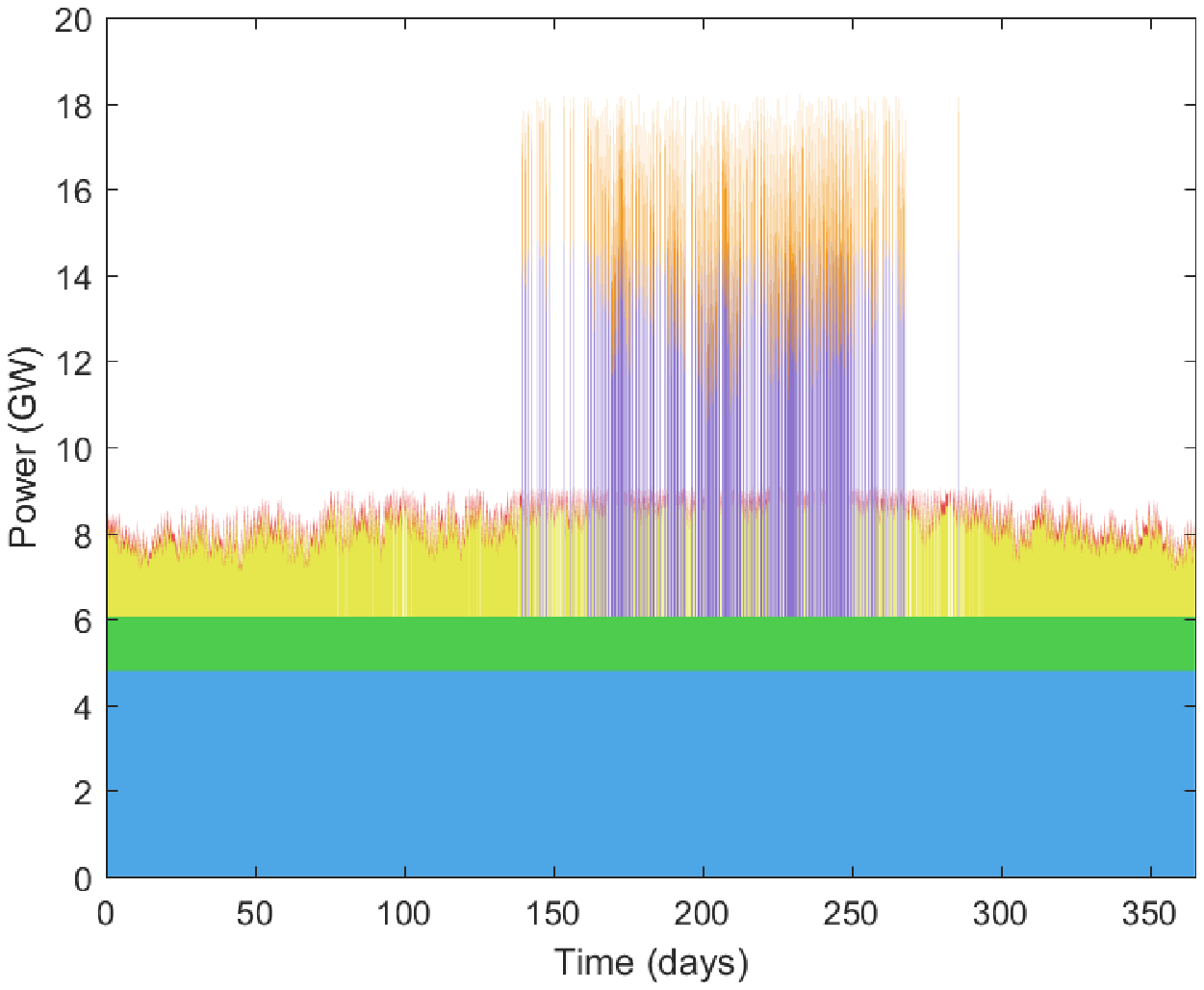}\\
    \includegraphics[width=.4\textwidth]{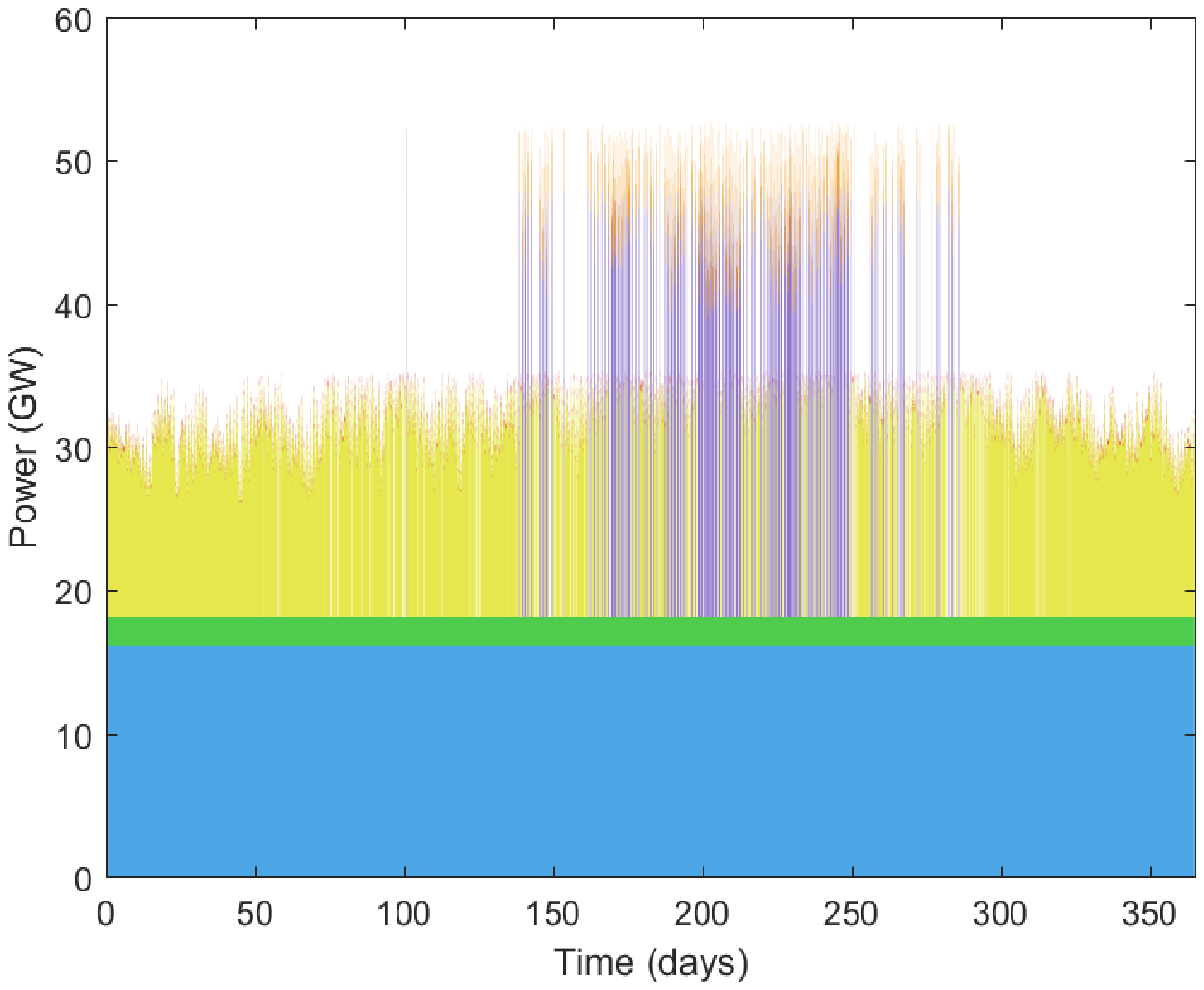}
  \end{tabular}
  }
\caption{Maximum charging power in North Atlantic, Continental and Mediterranean climate areas in 2019}
\label{fig:regs3}
\end{figure}

\begin{figure}
  \centering
  \resizebox{\ifdim\width>\columnwidth\columnwidth\else\width\fi}{!}{
  \begin{tabular}{c}
    \includegraphics[width=.4\textwidth]{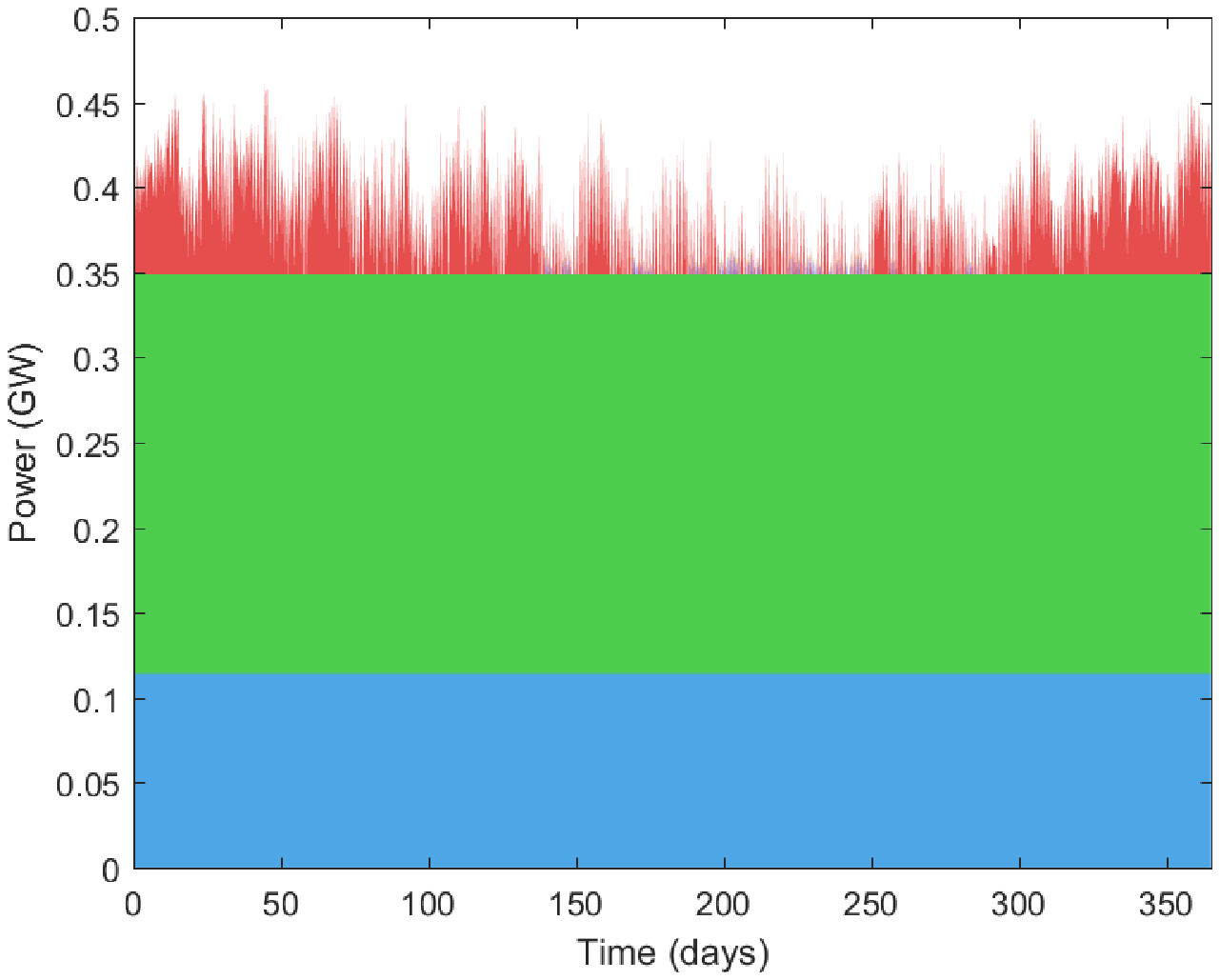}\\
  \includegraphics[width=.4\textwidth]{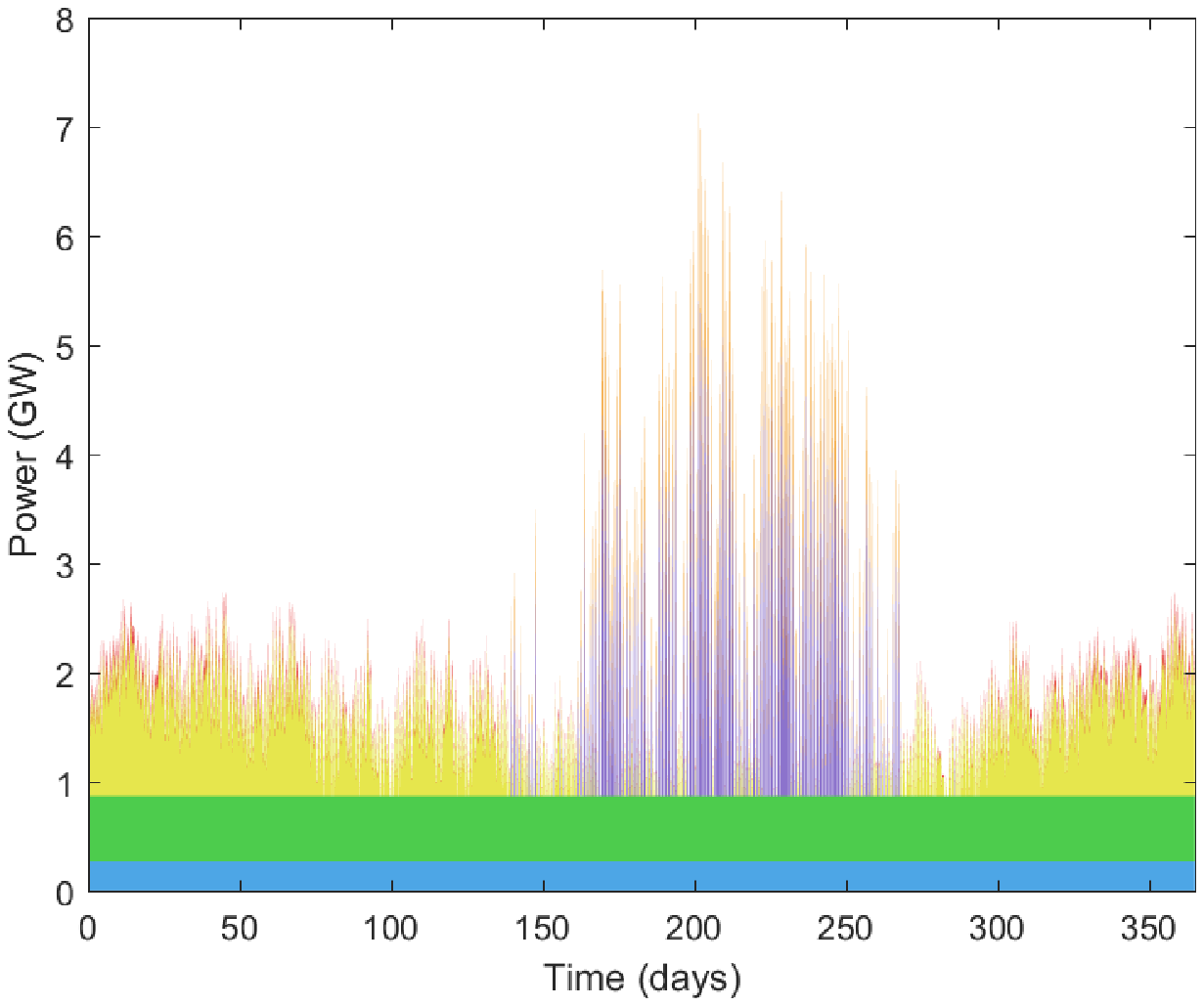}\\
    \includegraphics[width=.4\textwidth]{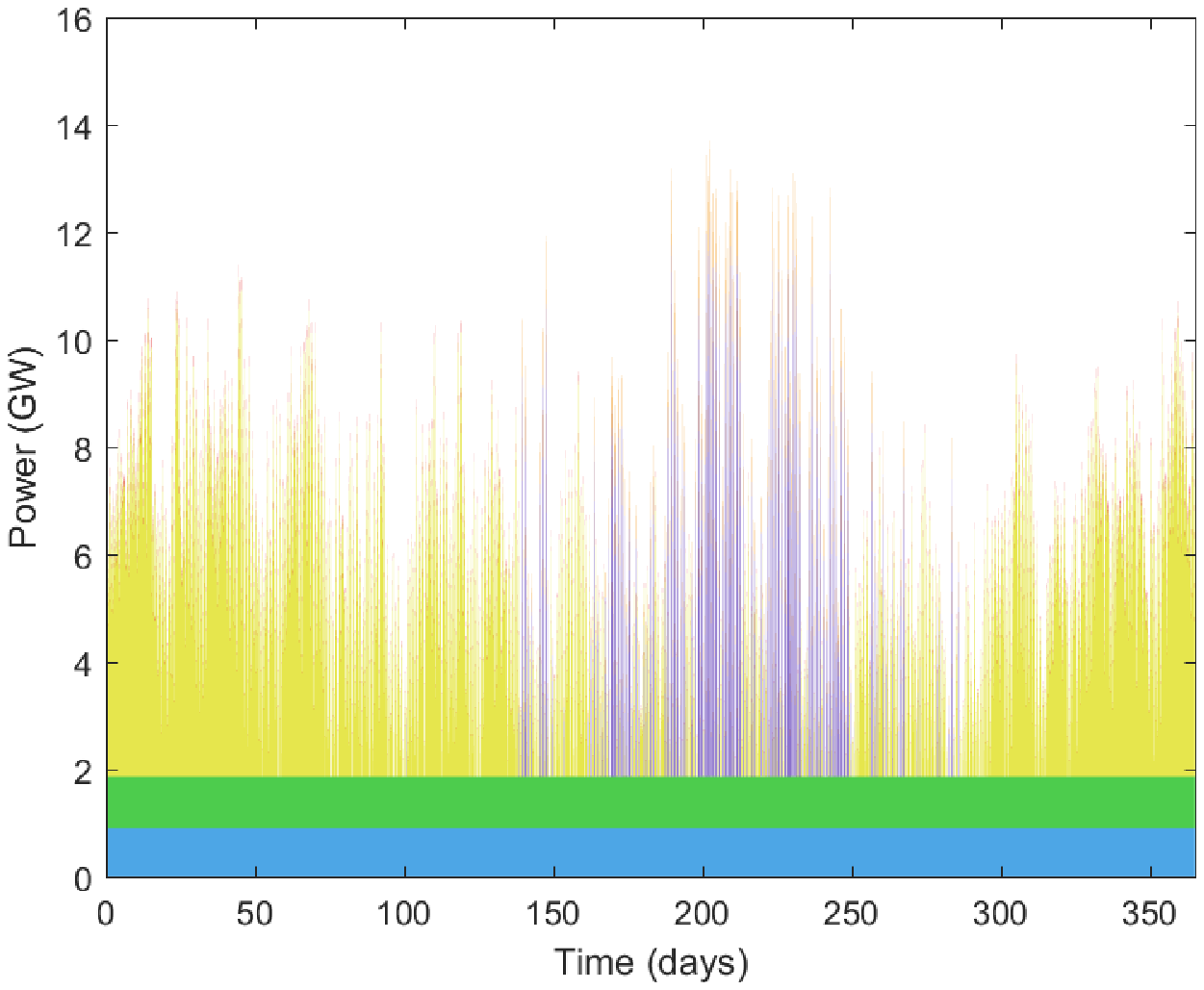}
   \end{tabular}
  }
\caption{Maximum discharging power in North Atlantic, Continental and Mediterranean climate areas in 2019}
\label{fig:regs4}
\end{figure}

The results of the study of the VB potential in Spain are reported here. 
Every existing TCL in each of the three climate areas of this country are
assumed be part of the corresponding VB. The charging capacity, discharging
capacity, maximum charging power, and maximum discharging power of each 
climate area of Spain throughout a natural year are shown in
Figures~\ref{fig:regs1}-\ref{fig:regs4}. 
The legend and colour codes for these plots are given in
Figure~\ref{fig:legend}. 
The ratios of greatest capacity and maximum power per home are shown in 
Table~\ref{ratios}.  
Information about the contribution of each residential type of TCL to the
VB potential of each climate region is collected in
Tables~\ref{ContributionChargeC}--\ref{ContributionDischargeP}. 
The average values from Table~\ref{typicalpara} for the parameters of each type
of TCL have been used in the study, along with the ambient temperature of
each climate area, obtained from the hourly temperature model~\cite{CTEhorario}
of the most populated cities in each climate area during 2019: 
Bilbao (North Atlantic), Madrid (Continental) and Barcelona (Mediterranean). 
In the case of refrigerators and electric water heaters, the ambient 
temperature is assumed to be constant and equal to 20${^\circ}$C.

\begin{table*}
\centering
\resizebox{\ifdim\width>\textwidth\textwidth\else\width\fi}{!}{
\begin{tabular}{cccc}
\toprule
 & \textbf{North Atlantic} & \textbf{Continental} & 
\textbf{Mediterranean} \\
\midrule
Greatest charging capacity/home (kWh) & 1.86 & 2.52 & 3.24 \\
Greatest discharging capacity/home (kWh)& 12.37 & 19.68 & 38.88 \\
Greatest maximum charging power/home (kW)& 1.13 & 2.93 & 5.29 \\
Greatest maximum discharging power/home (kW)& 0.19 & 1.18 & 1.39 \\
\bottomrule
\end{tabular}
}
\caption{Greatest capacity and maximum power ratios per climate area in 2019}
\label{ratios}
\end{table*}

\begin{table*}
\centering
\resizebox{\ifdim\width>\textwidth\textwidth\else\width\fi}{!}{
\begin{tabular}{cccc}
\toprule
\textbf{Charge capacity contribution (\%)} & \textbf{North Atlantic} & \textbf{Continental} & 
\textbf{Mediterranean} \\
\midrule
Water heater & 27.23 & 23.74 & 37.11 \\
Refrigerator & 72.16 & 69.14 & 51.16 \\
Reversible heat pump (cold) & 00.02 & 02.80 & 03.12 \\
Cold pump & 00.00 & 01.06 & 00.45 \\
Reversible heat pump (heat) & 00.00 & 02.93 & 07.96 \\
Non-reversible heat pump & 00.59 & 00.34 & 00.19 \\
\bottomrule
\end{tabular}
}
\caption{Contribution percentage of each type of TCL to charging capacity per climate area in 2019}
\label{ContributionChargeC}
\end{table*}

\begin{table*}
\centering
\resizebox{\ifdim\width>\textwidth\textwidth\else\width\fi}{!}{
\begin{tabular}{cccc}
\toprule
\textbf{Discharging capacity contribution (\%)} & \textbf{North Atlantic} & \textbf{Continental} & 
\textbf{Mediterranean} \\
\midrule
Water heater & 76.17 & 70.85 & 79.73 \\
Refrigerator & 23.34 & 23.85 & 12.71 \\
Reversible heat pump (cold) & 00.02 & 02.34 & 02.35 \\
Cold pump & 00.00 & 00.89 & 00.34 \\
Reversible heat pump (heat) & 00.00 & 01.85 & 04.76 \\
Non-reversible heat pump & 00.47 & 00.21 & 00.12 \\
\bottomrule
\end{tabular}
}
\caption{Contribution percentage of each type of TCL to discharging capacity per climate area in 2019}
\label{ContributionDishargeC}
\end{table*}

\begin{table*}
\centering
\resizebox{\ifdim\width>\textwidth\textwidth\else\width\fi}{!}{
\begin{tabular}{cccc}
\toprule
\textbf{Maximum charging power contribution (\%)} & \textbf{North Atlantic} & \textbf{Continental} & 
\textbf{Mediterranean} \\
\midrule
Water heater & 76.72 & 51.06 & 50.19 \\
Refrigerator & 18.34 & 13.41 & 06.24 \\
Reversible heat pump (cold) & 00.11 & 13.25 & 10.48 \\
Cold pump & 00.02 & 05.02 & 01.51 \\
Reversible heat pump (heat) & 00.00 & 15.47 & 30.83 \\
Non-reversible heat pump & 04.81 & 01.79 & 00.75 \\
\bottomrule
\end{tabular}
}
\caption{Contribution percentage of each type of TCL to maximum charging power per climate area in 2019}
\label{ContributionChargeP}
\end{table*}

\begin{table*}
\centering
\resizebox{\ifdim\width>\textwidth\textwidth\else\width\fi}{!}{
\begin{tabular}{cccc}
\toprule
\textbf{Maximum discharging power contribution (\%)} & \textbf{North Atlantic} & \textbf{Continental} & 
\textbf{Mediterranean} \\
\midrule
Water heater & 29.80 & 14.61 & 17.39 \\
Refrigerator & 61.29 & 33.01 & 18.60 \\
Reversible heat pump (cold) & 00.10 & 15.11 & 10.39 \\
Cold pump & 00.02 & 05.72 & 01.50 \\
Reversible heat pump (heat) & 00.00 & 28.28 & 50.88 \\
Non-reversible heat pump & 08.79 & 03.27 & 01.24 \\
\bottomrule
\end{tabular}
}
\caption{Contribution percentage of each type of TCL to maximum discharging power per climate area in 2019}
\label{ContributionDischargeP}
\end{table*}

Some interesting results should be highlighted. Mediterranean is the climate
area with the highest VB potential, while North Atlantic has the lowest. 
Electric water heaters and refrigerators have the highest contribution to the
charging and discharging capacity in the three climate areas. Moreover, this
contribution is constant during the year, because the ambient temperature
inside home is considered constant and equal to 20$^{\circ}$C.  Refrigerators
have more charging capacity in every climate area, while  water heaters have
more discharging capacity. These constant capacities can be used for demand management at any
time of the year. The remaining devices have a variable capacity contribution,
as can be easily seen by the numerous peaks and valleys appearing in
Figures~\ref{fig:regs1}--\ref{fig:regs4}. This is a consequence of the
variability of the temperature along different hours, days and seasons.
The capacity and maximum power of heat and cold pumps depend on the weather. 
In general, heat pumps can only be used on cold days and most of them are in
winter. In contrast, cold pumps are mostly used on hot days, typically 
during summer.
Reversible heat pumps and cold pumps have a greater contribution in Continental
and Mediterranean areas than in the North Atlantic area. 
The reason is the greater number of them in these areas (see Table
\ref{TCLnumber2017}). Consequently, the variability in capacity and maximum
power potential is more important in those climatic areas. Some summer days, 
the contribution of pumps to maximum charging and discharging power exceeds 
50\%. During these days, most of the cold pumps are working. 
Finally, the lack of symmetry in the charging and discharging process, 
explained in Section~\ref{VBco}, clearly manifests here as a greater
discharging capacity than charging capacity, as well as a greater
maximum charging power than maximum discharging power, both for every
climate area.

\subsection{Discussion} 

A huge amount of energy and power flexibility of TCLs can be efficiently
managed at real-time in a cost-effective way using a virtual battery. 
The control method proposed in this article to perform power regulation,
control, and communication between TCLs and the aggregator improves the 
accuracy in tracking the system operator command power signal and does not
require great investments in hardware or electronics because most of them are
already installed, including the TCLs, thermal sensors and Internet connection. 
This is an efficient and cost-effective alternative to conventional batteries
or fossil fuel solutions that require complete new installations. 
Furthermore, VB potential is expected to increase in the next few years
due to the progressive electrification of heating devices.

Another important benefit that VBs provide is the decentralization and
empowerment of consumers in the goal of balancing the grid. They become 
potential participants in balancing markets. Although these markets are not
currently completely developed in Europe, the European Union has launched
Regulation 2019/943 \cite{ReglamentoFlex} and Directive 2019/944
\cite{DirectivaFlex}, where balancing markets are considered. As mentioned in
\cite{ReglamentoFlex}, these markets can either be individually accessed or
be part of an aggregation, ensuring non-discriminatory access to all
participants and respecting the need to adapt to the increasing share of
variable generation and increased demand responsiveness. VBs fit properly with
these requirements.

In the case of Spain, net balancing energy amounted to approximately 1209~GWh in 2019
\cite{REEregulacion}. This means an average power of 138 MW. 
However, positive and negative hourly peaks of more than 4000~MW are
produced \cite{REEregulacion2}. The study performed in this paper shows that
taking advantage of FM trough VBs clearly helps to achieve power regulation goals in Spain, especially in extreme situations.
The remuneration for participating in balancing markets must be
important to encourage TCL owners to become contributors. Although the
Mediterranean climate area has the highest residential power and capacity
potential in Spain, VBs management should be profitable in any climate region
if the economic profit is sufficient.

\section{Conclusion} \label{conclusion}

The capacity and maximum power provided by the aggregation of TCLs in Spain
have been analysed in this paper. The different regions of Spain were 
classified into three different climate areas and a VB including an improved 
real-time control has been used for this study.
The obtained results show the flexibility potential of the different climate
areas of the country, which should encourage power system actors to develop and implement VBs technology.
In addition, a new control system for improving the operation of BV has been
developed. Its performance shows excellent levels of accuracy in tracking the
command signals provided by the system operator when VB constraints are
fulfilled. 

The results prove that VBs can be very useful in providing regulation support
to electrical grids. If the communication between TCLs and aggregators is 
sufficiently fast, VBs becomes a cost-effective and efficient short-term energy
storage alternative which, in combination with other existing ones (SMES,
flywheels, etc.), greatly facilitates renewable energy penetration.

VBs are a powerful instrument for implementing demand side 
management programs. Their utility is closely related to the speed and 
reliability of communication networks, the quality of TCLs models and 
the prediction of the ambient temperature. Improving these aspects is driving
our future research work.  

\vspace{6pt} 




\authorcontributions{Conceptualization, A.M-C., S.S-R. and E.B.; methodology, A.M-C. and S.S-R.; software, A.M-C.; validation, A.M-C., S.S-R. and E.B.; formal analysis, A.M-C.; investigation, A.M-C. and S.S-R.; resources, A.M-C. and S.S-R.; data curation, A.M-C.; writing---original draft preparation, A.M-C.; writing---review and editing, A.M-C., S.S-R. and E.B.; visualization, A.M-C. and E.B.; supervision, S.S-R. and E.B.; project administration, S.S-R.; funding acquisition, S.S-R. All authors have read and agreed to the published version of the manuscript.}


\funding{This research has been partially supported by the European Union ERDF (European
Regional Development Fund) and the Junta de Castilla y Le\'on
through ICE (Instituto para la Competitividad Empresarial) to improve 
innovation, technological development and research, dossier no. 
CCTT1/17/VA/0005.}

\conflictsofinterest{The authors declare no conflict of interest.} 



%

\appendixtitles{yes} 
\appendixstart
\appendix
%
%

\input{algorithms}

\end{paracol}
\reftitle{References}

\bibliography{base_bibliografia}

%


\end{document}

%% file: algorithms.tex
\section{Control Algorithms}

\begin{algorithm}[H]
\caption{Check of TCLs.}
\label{algoritmo1}
\begin{algorithmic}[1]

\STATE $P_{extra_{i}}^{k}=0$
\STATE $P_{-_{i}}=0$
\STATE $P_{+_{i}}=0$
\STATE Calculate $P_{0_{i}}^{k+1}$

\COMMENT {Total availability}
\IF {($\phi_{i} = 0$ and $\hat\theta_{a_{i}}^{k+1} < (\theta_{s_{i}} + \Delta_{i}$)) or ($\phi_{i} = 1$ and $\hat\theta_{a_{i}}^{k+1} > (\theta_{s_{i}} - \Delta_{i}$))}
	\STATE $\gamma_{i}^{k+1}=0$, $\delta_{i}^{k+1} =0$, $u_{i}^{k+1}=0$
		\IF {$\phi_{i} = 0$ and $\gamma_{i}^{k}=1$}
			\STATE $P_{extra_{i}}^{k} = P_{extra_{i}}^{k} + P_{0_{i}}^{k}$
		\ELSIF {$\phi_{i} = 1$ and $\gamma_{i}^{k}=1$}
			\STATE $P_{extra_{i}}^{k} = P_{extra_{i}}^{k} - P_{0_{i}}^{k}$
		\ENDIF
		\IF {$\phi_{i} = 0$ and $u_{i}^{k}=1$}
			\STATE $P_{extra_{i}}^{k} = P_{extra_{i}}^{k} - P_{i}$
			\STATE $\zeta_{i}^{k+1}=0$
		\ELSIF {$\phi_{i} = 1$ and $u_{i}^{k}=1$}
			\STATE $P_{extra_{i}}^{k} = P_{extra_{i}}^{k} + P_{i}$
			\STATE $\zeta_{i}^{k+1}=0$
		\ENDIF
\ENDIF

\IF {$\gamma_{i}^{k}=0$ and $\gamma_{i}^{k+1}=1$}
	\IF {$\phi_{i} = 0$}
		\STATE $P_{extra_{i}}^{k} = P_{extra_{i}}^{k} - P_{0_{i}}^{k+1}$
	\ELSIF {$\phi_{i} = 1$}
		\STATE $P_{extra_{i}}^{k} = P_{extra_{i}}^{k} + P_{0_{i}}^{k+1}$
	\ENDIF
\ENDIF

\COMMENT {Forecast ambient temperature}
\IF {$\gamma_{i}^{k}=1$ and $\gamma_{i}^{k+1}=1$}
	\IF {$\phi_{i} = 0$}
		\STATE $P_{extra_{i}}^{k} = P_{extra_{i}}^{k} - P_{0_{i}}^{k+1} + P_{0_{i}}^{k}$
	\ELSIF {$\phi_{i} = 1$}
		\STATE $P_{extra_{i}}^{k} = P_{extra_{i}}^{k} + P_{0_{i}}^{k+1}- P_{0_{i}}^{k}$
	\ENDIF
\ENDIF

\COMMENT {TCL temperature}
\STATE Calculate $\theta_{i}^{k+1}$
\IF {$\gamma_{i}^{k+1}=1$}
	\IF {$\theta_{i}^{k+1} \geq (\theta_{s_{i}} + \Delta_{i})$}
		\STATE $\delta_{i}^{k+1} =0$
		\IF {$\phi_{i} = 0$}
			\STATE $P_{-_{i}}^{k} = P_{-_{i}}^{k}+ P_{i}$
			\IF {$u_{i}^{k} = 0$}
				\STATE $u_{i}^{k+1} = 1$
				\STATE $P_{extra_{i}}^{k} = P_{extra_{i}}^{k} + P_{i}$
				\STATE $\zeta_{i}^{k+1} = 0$
			\ENDIF
			
			\algstore{myalg}
			\end{algorithmic}
			\end{algorithm}

			\begin{algorithm}                    
			\begin{algorithmic} [1]
			\algrestore{myalg}
			
		\ELSIF {$\phi_{i} = 1$}
			\STATE $P_{+_{i}}^{k} = P_{+_{i}}^{k} - P_{i}$
			\IF {$u_{i}^{k} = 1$}
				\STATE $u_{i}^{k+1} = 0$
				\STATE $P_{extra_{i}}^{k} = P_{extra_{i}}^{k} + P_{i}$
				\STATE $\zeta_{i}^{k+1} = 0$
			\ENDIF
		\ENDIF
	\ELSIF {$\theta_{i}^{k+1} \leq (\theta_{s_{i}} - \Delta_{i})$}
		\STATE $\delta_{i}^{k+1} = 0$
		\IF {$\phi_{i} = 0$}
			\STATE $P_{+_{i}}^{k} = P_{+_{i}}^{k}+ P_{i}$
			\IF {$u_{i}^{k} = 1$}
				\STATE $u_{i}^{k+1} = 0$
				\STATE $P_{extra_{i}}^{k} = P_{extra_{i}}^{k} - P_{i}$
				\STATE $\zeta_{i}^{k+1} = 0$
			\ENDIF
		\ELSIF {$\phi_{i} = 1$}
			\STATE $P_{-_{i}}^{k} = P_{-_{i}}^{k} - P_{i}$
			\IF {$u_{i}^{k} = 0$}
				\STATE $u_{i}^{k+1} = 1$
				\STATE $P_{extra_{i}}^{k} = P_{extra_{i}}^{k} - P_{i}$
				\STATE $\zeta_{i}^{k+1} = 0$
			\ENDIF
		\ENDIF
	\ENDIF
\ENDIF

\COMMENT {Cycled passed time}
\IF {$\zeta_{i}^{k+1} \leq \kappa_{i}$}
	\STATE $\delta_{i}^{k+1} = 0$
	\IF {$u_{i}^{k+1} = 1$ and $\theta_{i}^{k+1} \leq \theta_{s_{i}} + \Delta_{i}$ and $\theta_{i}^{k+1} \geq \theta_{s_{i}} - \Delta_{i}$ and $\zeta_{i}^{k+1} \neq 0$ and $\gamma_{i}^{k+1}=1$}
		\IF {$\phi_{i} = 0$}
			\STATE $P_{-_{i}}^{k} = P_{-_{i}}^{k} + P_{i}$
		\ELSIF {$\phi_{i} = 1$}
			\STATE $P_{-_{i}}^{k} = P_{-_{i}}^{k} - P_{i}$
		\ENDIF
	\ELSIF {$u_{i}^{k+1} = 0$ and $\theta_{i}^{k+1} \leq \theta_{s_{i}} + \Delta_{i}$ and $\theta_{i}^{k+1} \geq \theta_{s_{i}} - \Delta_{i}$ and $\zeta_{i}^{k+1} \neq 0$ and $\gamma_{i}^{k+1}=1$}
		\IF {$\phi_{i} = 0$}
			\STATE $P_{+_{i}}^{k} = P_{+_{i}}^{k} + P_{i}$
		\ELSIF {$\phi_{i} = 1$}
			\STATE $P_{+_{i}}^{k} = P_{+_{i}}^{k} - P_{i}$
		\ENDIF
	\ENDIF
\ENDIF

\end{algorithmic}
\end{algorithm}

\begin{algorithm}
\begin{algorithmic}[1]

\STATE $P_{extra}^{k}=0$

\FOR {$i$ from 1 to $N$}
	\STATE $P_{extra}^{k} = P_{extra}^{k} + P_{extra_{i}}^{k}$
	\STATE $P_{+}^{k} = P_{+}^{k} + P_{+_{i}}^{k}$
	\STATE $P_{-}^{k} = P_{-}^{k} + P_{-_{i}}^{k}$
\ENDFOR

\COMMENT {Capacity}
\STATE Calculate $C_{c}^{k+1}$, $C_{d}^{k+1}$

\COMMENT {State of charge}
\STATE Calculate $SOC_{c}^{k+1}$, $SOC_{d}^{k+1}$

\COMMENT {Maximum power}
\STATE Calculate $n_{+}^{k+1}$, $n_{-}^{k+1}$, $n_{+}^{'k+1}$, $n_{-}^{'k+1}$

\COMMENT {Regulation signal}
\STATE Calculate $P_{agg}^{k}$, $P_{base}^{k}$, $\psi^{k}$, $\epsilon^{k}$

\end{algorithmic}
\caption{Aggregation.}
\label{algoritmo2}
\end{algorithm}

\begin{algorithm}
\begin{algorithmic}[1]

\STATE $p^{k}=0$

\COMMENT {Disaggregation}
\IF {$\epsilon^k$ $<$ 0}
	\STATE Calculate $\lambda_{i}^k$
	\STATE Order TCL from bigger $\lambda_{i}^k$ to smaller $\lambda_{i}^k$
	\WHILE{$|\epsilon^k|$ $>$ $|p^k|$ and $i$ $<$ $N$}
		\STATE $i$ = $i$ + 1
		\IF {$u^k$ = 1 and $\delta^{k+1}$ = 1}
			\STATE $u^{k+1}$ = 0
			\STATE $\zeta^{k+1}$ = 0
			\IF {$\phi = 0$}
				\STATE $p^{k}$ = $p^{k}$ + $P_i$
				\STATE $P_{+}^{k+1}$ = $P_{+}^{k+1}$ + $P_i$
			\ELSIF {$\phi = 1$}
				\STATE $p^{k}$ = $p^{k}$ - $P_i$
				\STATE $P_{+}^{k+1}$ = $P_{+}^{k+1}$ - $P_i$
			\ENDIF
		\ENDIF
	\ENDWHILE
\ELSIF {$\epsilon^k$ $>$ 0}
	\STATE Calculate $\lambda_{i}^k$
	\STATE Order TCL from bigger $\lambda_{i}^k$ to smaller $\lambda_{i}^k$
	\WHILE{$|\epsilon^k|$ $>$ $|p^k|$ and $i$ $<$ $N$}
		\STATE $i$ = $i$ + 1
		\IF {$u^k$ = 0 and $\delta^{k+1}$ = 1}
			\STATE $u^{k+1}$ = 1
			\STATE $\zeta^{k+1}$ = 0
			\IF {$\phi = 0$}
				\STATE $p^{k}$ = $p^{k}$ + $P_i$
				\STATE $P_{-}^{k+1}$ = $P_{-}^{k+1}$ + $P_i$
			\ELSIF {$\phi = 1$}
				\STATE $p^{k}$ = $p^{k}$ - $P_i$
				\STATE $P_{-}^{k+1}$ = $P_{-}^{k+1}$ - $P_i$
			\ENDIF
		\ENDIF
	\ENDWHILE
\ENDIF

\end{algorithmic}
\caption{Priority Control.}
\label{algoritmo3}
\end{algorithm}

%% file: Articulo.bbl
\begin{thebibliography}{999}

\bibitem[{Callaway} and {Hiskens}(2011)]{Callaway2011achieving}
{Callaway}, D.S.; {Hiskens}, I.A.
\newblock Achieving Controllability of Electric Loads.
\newblock {\em Proceedings of the IEEE} {\bf 2011}, {\em 99},~184--199.
\newblock
  doi:{\changeurlcolor{black}\href{https://doi.org/10.1109/JPROC.2010.2081652}{\detokenize{10.1109/JPROC.2010.2081652}}}.

\bibitem[Koch \em{et~al.}(2011)Koch, Mathieu, and Callaway]{koch2011modeling}
Koch, S.; Mathieu, J.L.; Callaway, D.S.
\newblock Modeling and control of aggregated heterogeneous thermostatically
  controlled loads for ancillary services.
\newblock  Proc. PSCC. Citeseer,  2011, pp. 1--7.

\bibitem[Hao \em{et~al.}(2013)Hao, Sanandaji, Poolla, and Vincent]{HHao1}
Hao, H.; Sanandaji, B.M.; Poolla, K.; Vincent, T.L.
\newblock A generalized battery model of a collection of
  \uppercase{t}hermostatically \uppercase{c}ontrolled \uppercase{l}oads for
  providing ancillary service.
\newblock  2013 51st Annual Allerton Conference on Communication, Control, and
  Computing (Allerton),  2013, pp. 551--558.
\newblock
  doi:{\changeurlcolor{black}\href{https://doi.org/10.1109/Allerton.2013.6736573}{\detokenize{10.1109/Allerton.2013.6736573}}}.

\bibitem[Acharya \em{et~al.}(2017)Acharya, El~Moursi, and
  Al-Hinai]{acharya2017coordinated}
Acharya, S.; El~Moursi, M.S.; Al-Hinai, A.
\newblock Coordinated frequency control strategy for an islanded microgrid with
  demand side management capability.
\newblock {\em IEEE Transactions on Energy Conversion} {\bf 2017}, {\em
  33},~639--651.

\bibitem[Tindemans \em{et~al.}(2015)Tindemans, Trovato, and
  Strbac]{tindemans2015decentralized}
Tindemans, S.H.; Trovato, V.; Strbac, G.
\newblock Decentralized control of thermostatic loads for flexible demand
  response.
\newblock {\em IEEE Transactions on Control Systems Technology} {\bf 2015},
  {\em 23},~1685--1700.

\bibitem[Wang \em{et~al.}(2020)Wang, Wu, and Kalsi]{wang2020flexibility}
Wang, P.; Wu, D.; Kalsi, K.
\newblock Flexibility Estimation and Control of Thermostatically Controlled
  Loads with Lock Time for Regulation Service.
\newblock {\em IEEE Transactions on Smart Grid} {\bf 2020}.

\bibitem[{Mathieu} \em{et~al.}(2014){Mathieu}, {Rasmussen}, {S{\o}rensen},
  {J{\`o}hannsson}, and {Andersson}]{6938939}
{Mathieu}, J.L.; {Rasmussen}, T.B.; {S{\o}rensen}, M.; {J{\`o}hannsson}, H.;
  {Andersson}, G.
\newblock Technical resource potential of non-disruptive residential demand
  response in \uppercase{d}enmark.
\newblock  2014 IEEE PES General Meeting | Conference Exposition,  2014, pp.
  1--5.
\newblock
  doi:{\changeurlcolor{black}\href{https://doi.org/10.1109/PESGM.2014.6938939}{\detokenize{10.1109/PESGM.2014.6938939}}}.

\bibitem[Gils(2016)]{GILS2016401}
Gils, H.C.
\newblock Economic potential for future demand response in Germany {\textminus}
  Modeling approach and case study.
\newblock {\em Applied Energy} {\bf 2016}, {\em 162},~401 -- 415.
\newblock
  doi:{\changeurlcolor{black}\href{https://doi.org/https://doi.org/10.1016/j.apenergy.2015.10.083}{\detokenize{https://doi.org/10.1016/j.apenergy.2015.10.083}}}.

\bibitem[Trovato \em{et~al.}(2016)Trovato, Sanz, Chaudhuri, and
  Strbac]{trovato2016advanced}
Trovato, V.; Sanz, I.M.; Chaudhuri, B.; Strbac, G.
\newblock Advanced control of thermostatic loads for rapid frequency response
  in Great Britain.
\newblock {\em IEEE Transactions on Power Systems} {\bf 2016}, {\em
  32},~2106--2117.

\bibitem[Kamgarpour \em{et~al.}(2014)Kamgarpour, Vrettos, Andersson, and
  Lygeros]{kamgarpour2014population}
Kamgarpour, M.; Vrettos, E.; Andersson, G.; Lygeros, J.
\newblock Population of \uppercase{t}hermostatically \uppercase{c}ontrolled
  \uppercase{l}oads for the \uppercase{s}wiss \uppercase{a}ncillary
  \uppercase{s}ervice \uppercase{m}arket.
\newblock  Proceedings of the COLEB 2014 Workshop Computational Optimisation of
  Low-Energy Buildings. 6 \& 7 March 2014, ETH Z{\"u}rich. Citeseer,  2014, pp.
  49--50.

\bibitem[Hao \em{et~al.}(2015)Hao, Sanandaji, Poolla, and
  Vincent]{hao2015potentials}
Hao, H.; Sanandaji, B.M.; Poolla, K.; Vincent, T.L.
\newblock Potentials and economics of residential thermal loads providing
  regulation reserve.
\newblock {\em Energy Policy} {\bf 2015}, {\em 79},~115--126.

\bibitem[Conte \em{et~al.}(2017)Conte, Massucco, Silvestro, Ciapessoni, and
  Cirio]{conte2017stochastic}
Conte, F.; Massucco, S.; Silvestro, F.; Ciapessoni, E.; Cirio, D.
\newblock Stochastic modelling of aggregated thermal loads for impact analysis
  of demand side frequency regulation in the case of \uppercase{s}ardinia in
  2020.
\newblock {\em International Journal of Electrical Power \& Energy Systems}
  {\bf 2017}, {\em 93},~291--307.

\bibitem[{Instituto para la {Diversificaci\'on} y ahorro de la
  {Energ\'ia}}(2011)]{sech2011analisis}
{Instituto para la {Diversificaci\'on} y ahorro de la {Energ\'ia}}.
\newblock Proyecto \uppercase{shec-spahousec}. \uppercase{a}n{\'a}lisis del
  consumo energ{\'e}tico del sector residencial en \uppercase{e}spa{\~n}a.
  \uppercase{i}nforme final.
\newblock
  \url{http://www.idae.es/uploads/documentos/documentos_Informe_SPAHOUSEC_ACC_f68291a3.pdf},
   2011.
\newblock [Last accessed date: 2021-02-19].

\bibitem[Han \em{et~al.}(2011)Han, Han, and Sezaki]{han2011estimation}
Han, S.; Han, S.; Sezaki, K.
\newblock Estimation of achievable power capacity from plug-in electric
  vehicles for V2G frequency regulation: Case studies for market participation.
\newblock {\em IEEE Transactions on Smart Grid} {\bf 2011}, {\em 2},~632--641.

\bibitem[Liu \em{et~al.}(2013)Liu, Hu, Song, and Lin]{liu2013decentralized}
Liu, H.; Hu, Z.; Song, Y.; Lin, J.
\newblock Decentralized vehicle-to-grid control for primary frequency
  regulation considering charging demands.
\newblock {\em IEEE Transactions on Power Systems} {\bf 2013}, {\em
  28},~3480--3489.

\bibitem[Kempton \em{et~al.}(2008)Kempton, Udo, Huber, Komara, Letendre, Baker,
  Brunner, and Pearre]{kempton2008test}
Kempton, W.; Udo, V.; Huber, K.; Komara, K.; Letendre, S.; Baker, S.; Brunner,
  D.; Pearre, N.
\newblock A test of vehicle-to-grid (V2G) for energy storage and frequency
  regulation in the PJM system.
\newblock {\em Results from an Industry-University Research Partnership} {\bf
  2008}, {\em 32}.

\bibitem[Hao \em{et~al.}(2015)Hao, Sanandaji, Poolla, and Vincent]{HHao2}
Hao, H.; Sanandaji, B.M.; Poolla, K.; Vincent, T.L.
\newblock Aggregate \uppercase{f}lexibility of \uppercase{t}hermostatically
  \uppercase{c}ontrolled \uppercase{l}oads.
\newblock {\em IEEE Transactions on Power Systems} {\bf 2015}, {\em
  30},~189--198.
\newblock
  doi:{\changeurlcolor{black}\href{https://doi.org/10.1109/TPWRS.2014.2328865}{\detokenize{10.1109/TPWRS.2014.2328865}}}.

\bibitem[Nandanoori \em{et~al.}(2019)Nandanoori, Chakraborty, Ramachandran, and
  Kundu]{nandanoori2019identification}
Nandanoori, S.P.; Chakraborty, I.; Ramachandran, T.; Kundu, S.
\newblock {Identification and Validation of Virtual Battery Model for
  Heterogeneous Devices}.
\newblock  Proceedings of the IEEE Power \& Energy Society General Meeting
  (PESGM 2019),  2019, pp. 1--5.
\newblock
  doi:{\changeurlcolor{black}\href{https://doi.org/10.1109/PESGM40551.2019.8973978}{\detokenize{10.1109/PESGM40551.2019.8973978}}}.

\bibitem[Ding \em{et~al.}(2019)Ding, Cui, Zhang, Hui, Qiu, and
  Song]{DING201946}
Ding, Y.; Cui, W.; Zhang, S.; Hui, H.; Qiu, Y.; Song, Y.
\newblock Multi-state operating reserve model of aggregate
  thermostatically-controlled-loads for power system short-term reliability
  evaluation.
\newblock {\em Applied Energy} {\bf 2019}, {\em 241},~46 -- 58.

\bibitem[Lu \em{et~al.}(2005)Lu, Chassin, and Widergren]{lu2005modeling}
Lu, N.; Chassin, D.P.; Widergren, S.E.
\newblock Modeling uncertainties in aggregated thermostatically controlled
  loads using a state queueing model.
\newblock {\em IEEE Transactions on Power Systems} {\bf 2005}, {\em
  20},~725--733.

\bibitem[Zhang \em{et~al.}(2012)Zhang, Kalsi, Fuller, Elizondo, and
  Chassin]{zhang2012aggregate}
Zhang, W.; Kalsi, K.; Fuller, J.; Elizondo, M.; Chassin, D.
\newblock Aggregate model for heterogeneous thermostatically controlled loads
  with demand response.
\newblock  2012 IEEE Power and Energy Society General Meeting. IEEE,  2012, pp.
  1--8.

\bibitem[Perfumo \em{et~al.}(2012)Perfumo, Kofman, Braslavsky, and
  Ward]{perfumo2012load}
Perfumo, C.; Kofman, E.; Braslavsky, J.H.; Ward, J.K.
\newblock Load management: \uppercase{m}odel-based control of aggregate power
  for populations of thermostatically controlled loads.
\newblock {\em Energy Conversion and Management} {\bf 2012}, {\em 55},~36--48.

\bibitem[Kundu \em{et~al.}(2011)Kundu, Sinitsyn, Backhaus, and
  Hiskens]{kundu2011modeling}
Kundu, S.; Sinitsyn, N.; Backhaus, S.; Hiskens, I.
\newblock Modeling and control of thermostatically controlled loads.
\newblock {\em arXiv preprint arXiv:1101.2157} {\bf 2011}.

\bibitem[{Radaideh} \em{et~al.}(2019){Radaideh}, {Vaidya}, and
  {Ajjarapu}]{7995082}
{Radaideh}, A.; {Vaidya}, U.; {Ajjarapu}, V.
\newblock \uppercase{s}equential \uppercase{s}et-\uppercase{p}oint
  \uppercase{c}ontrol for \uppercase{h}eterogeneous
  \uppercase{t}hermostatically \uppercase{c}ontrolled \uppercase{l}oads
  \uppercase{t}hrough an \uppercase{e}xtended \uppercase{m}arkov
  \uppercase{c}hain Abstraction.
\newblock {\em IEEE Transactions on Smart Grid} {\bf 2019}, {\em 10},~116--127.
\newblock
  doi:{\changeurlcolor{black}\href{https://doi.org/10.1109/TSG.2017.2732949}{\detokenize{10.1109/TSG.2017.2732949}}}.

\bibitem[Mathieu and Callaway(2012)]{mathieu2012state}
Mathieu, J.L.; Callaway, D.S.
\newblock State estimation and control of heterogeneous thermostatically
  controlled loads for load following.
\newblock  2012 45th Hawaii International Conference on System Sciences. IEEE,
  2012, pp. 2002--2011.

\bibitem[Xia \em{et~al.}(2019)Xia, Song, and Chen]{xia2019hierarchical}
Xia, M.; Song, Y.; Chen, Q.
\newblock Hierarchical control of thermostatically controlled loads oriented
  smart buildings.
\newblock {\em Applied Energy} {\bf 2019}, {\em 254},~113493.

\bibitem[Liu and Shi(2015)]{liu2015model}
Liu, M.; Shi, Y.
\newblock Model predictive control of aggregated heterogeneous second-order
  thermostatically controlled loads for ancillary services.
\newblock {\em IEEE transactions on power systems} {\bf 2015}, {\em
  31},~1963--1971.

\bibitem[Liu \em{et~al.}(2019)Liu, Peeters, Callaway, and
  Claessens]{liu2019trajectory}
Liu, M.; Peeters, S.; Callaway, D.S.; Claessens, B.J.
\newblock Trajectory tracking with an aggregation of domestic hot water
  heaters: \uppercase{c}ombining model-based and model-free control in a
  commercial deployment.
\newblock {\em IEEE Transactions on Smart Grid} {\bf 2019}.

\bibitem[Khan \em{et~al.}(2016)Khan, Shahzad, Habib, Gawlik, and
  Palensky]{SKhan}
Khan, S.; Shahzad, M.; Habib, U.; Gawlik, W.; Palensky, P.
\newblock Stochastic battery model for aggregation of thermostatically
  controlled loads.
\newblock  2016 IEEE International Conference on Industrial Technology (ICIT),
  2016, pp. 570--575.
\newblock
  doi:{\changeurlcolor{black}\href{https://doi.org/10.1109/ICIT.2016.7474812}{\detokenize{10.1109/ICIT.2016.7474812}}}.

\bibitem[{Conte} \em{et~al.}(2018){Conte}, {di Vergagni}, {Massucco},
  {Silvestro}, {Ciappessoni}, and {Cirio}]{8586247}
{Conte}, F.; {di Vergagni}, M.C.; {Massucco}, S.; {Silvestro}, F.;
  {Ciappessoni}, E.; {Cirio}, D.
\newblock Frequency \uppercase{r}egulation by \uppercase{t}hermostatically
  \uppercase{c}ontrolled \uppercase{l}oads: a \uppercase{t}echnical and
  \uppercase{e}conomic \uppercase{a}nalysis.
\newblock  2018 IEEE Power Energy Society General Meeting (PESGM),  2018, pp.
  1--5.
\newblock
  doi:{\changeurlcolor{black}\href{https://doi.org/10.1109/PESGM.2018.8586247}{\detokenize{10.1109/PESGM.2018.8586247}}}.

\bibitem[Bogodorova \em{et~al.}(2016)Bogodorova, Vanfretti, and
  Turitsyn]{bogodorova2016voltage}
Bogodorova, T.; Vanfretti, L.; Turitsyn, K.
\newblock Voltage control-based ancillary service using thermostatically
  controlled loads.
\newblock  2016 IEEE Power and Energy Society General Meeting (PESGM). IEEE,
  2016, pp. 1--5.

\bibitem[{Wang} \em{et~al.}(2019){Wang}, {Tang}, {Xu}, and {Xu}]{8606271}
{Wang}, Y.; {Tang}, Y.; {Xu}, Y.; {Xu}, Y.
\newblock A \uppercase{d}istributed \uppercase{c}ontrol \uppercase{s}cheme of
  \uppercase{t}hermostatically \uppercase{c}ontrolled \uppercase{l}oads for the
  \uppercase{b}uilding-\uppercase{m}icrogrid \uppercase{c}ommunity.
\newblock {\em IEEE Transactions on Sustainable Energy} {\bf 2019}, pp. 1--1.
\newblock
  doi:{\changeurlcolor{black}\href{https://doi.org/10.1109/TSTE.2019.2891072}{\detokenize{10.1109/TSTE.2019.2891072}}}.

\bibitem[Cheng \em{et~al.}(2017)Cheng, Sami, and Wu]{cheng2017benefits}
Cheng, M.; Sami, S.S.; Wu, J.
\newblock Benefits of using virtual energy storage system for power system
  frequency response.
\newblock {\em Applied energy} {\bf 2017}, {\em 194},~376--385.

\bibitem[Mathieu \em{et~al.}(2014)Mathieu, Kamgarpour, Lygeros, Andersson, and
  Callaway]{mathieu2014arbitraging}
Mathieu, J.L.; Kamgarpour, M.; Lygeros, J.; Andersson, G.; Callaway, D.S.
\newblock Arbitraging intraday wholesale energy market prices with aggregations
  of thermostatic loads.
\newblock {\em IEEE Transactions on Power Systems} {\bf 2014}, {\em
  30},~763--772.

\bibitem[{Mart\'in Crespo} and {Saludes Rodil}(2018)]{Yo2018}
{Mart\'in Crespo}, A.; {Saludes Rodil}, S.
\newblock Uso de la flexibilidad de cargas el\'ectricas con inercia t\'ermica
  mediante una bater\'ia virtual para la regulaci\'on de red.
\newblock {\em V Congreso Smart Grids: Libro de Comunicaciones} {\bf 2018}.

\bibitem[Mathieu(2012)]{osti_1182734}
Mathieu, J.L.
\newblock Modeling, analysis, and control of demand response resources.
\newblock PhD thesis, UC Berkeley,  2012.

\bibitem[Lakshmanan \em{et~al.}(2016)Lakshmanan, Marinelli, Kosek,
  N{\o}rg{\aa}rd, and Bindner]{LAKSHMANAN2016705}
Lakshmanan, V.; Marinelli, M.; Kosek, A.M.; N{\o}rg{\aa}rd, P.B.; Bindner, H.W.
\newblock Impact of thermostatically controlled loads' demand response
  activation on aggregated power: \uppercase{a} field experiment.
\newblock {\em Energy} {\bf 2016}, {\em 94},~705 -- 714.

\bibitem[{Instituto {N}acional de {E}stad\'istica}()]{INEENI}
{Instituto {N}acional de {E}stad\'istica}.
\newblock N\'umero de hogares por provincias seg\'un tipo de hogar y n\'umero
  de habitaciones de la vivienda.
\newblock
  \url{http://www.ine.es/jaxi/Tabla.htm?path=/t20/p274/serie/def/p03/l0/&file=03003.px&L=0}.
\newblock [Last accessed date: 2021-02-19].

\bibitem[Mat(2019)]{MATLAB:2019b}
The Mathworks, Inc., Natick, Massachusetts.
\newblock {\em {MATLAB version 2019b}},  2019.

\bibitem[{Espa\~na}(2016)]{BOEref}
{Espa\~na}.
\newblock {Resoluci\'on de 13 de julio de 2006, de la \uppercase{s}ecretar\'ia
  \uppercase{g}eneral de \uppercase{e}nerg\'ia, por la que se aprueba el
  procedimiento de operaci\'on 1.5 \guillemotleft \uppercase{e}stablecimiento
  de la reserva para la regulaci\'on frecuencia-potencia\guillemotright}.
\newblock {\em Bolet\'in Oficial del Estado} {\bf 2016}, {\em
  173},~27473--27474.

\bibitem[{Smart Energy Demand Coalition}()]{SEDCp}
{Smart Energy Demand Coalition}.
\newblock Explicit \uppercase{d}emand \uppercase{r}esponse \uppercase{i}n
  \uppercase{e}urope {\textminus} \uppercase{m}apping the \uppercase{m}arkets
  2017.
\newblock
  \url{https://www.smarten.eu/wp-content/uploads/2017/04/SEDC-Explicit-Demand-Response-in-Europe-Mapping-the-Markets-2017.pdf}.
\newblock [Last accessed date: 2021-02-19].

\bibitem[{Espa\~na}(2017)]{CTEhorario}
{Espa\~na}.
\newblock \uppercase{d}ocumento \uppercase{b}\'asico \uppercase{he}
  \uppercase{a}horro de \uppercase{e}nerg\'ia.
\newblock {\em \uppercase{c}\'odigo \uppercase{t}\'ecnico de la
  \uppercase{e}dificaci\'on} {\bf 2017}.

\bibitem[{European Union}(2019{\natexlab{a}})]{ReglamentoFlex}
{European Union}.
\newblock {Regulation (\uppercase{eu}) 2019/943 of the \uppercase{e}uropean
  \uppercase{p}arliament and of the \uppercase{c}ouncil of 5 \uppercase{j}une
  2019 on the internal market for electricity}.
\newblock {\em Official \uppercase{j}ournal of the \uppercase{e}uropean
  \uppercase{u}nion} {\bf 2019}, {\em L 158/154}.

\bibitem[{European Union}(2019{\natexlab{b}})]{DirectivaFlex}
{European Union}.
\newblock {Directive (\uppercase{eu}) 2019/944 of the \uppercase{e}uropean
  \uppercase{p}arliament and of the \uppercase{c}ouncil of 5 \uppercase{j}une
  2019 on common rules for the internal market for electricity and amending
  \uppercase{d}irective 2012/27/\uppercase{eu}}.
\newblock {\em Official \uppercase{j}ournal of the \uppercase{e}uropean
  \uppercase{u}nion} {\bf 2019}, {\em L 158/125}.

\bibitem[{Red El\'ectrica de Espa\~na}({\natexlab{a}})]{REEregulacion}
{Red El\'ectrica de Espa\~na}.
\newblock {Net balancing energy}.
\newblock
  \url{https://www.esios.ree.es/en/analysis/762?vis=1&start_date=01-01-2019T00%3A00&end_date=31-12-2019T23%3A50&compare_start_date=01-01-2018T00%3A00&groupby=year}.
\newblock [Last accessed date: 2021-02-19].

\bibitem[{Red El\'ectrica de Espa\~na}({\natexlab{b}})]{REEregulacion2}
{Red El\'ectrica de Espa\~na}.
\newblock {Net balancing energy}.
\newblock
  \url{https://www.esios.ree.es/en/analysis/762?vis=1&start_date=01-01-2019T00%3A00&end_date=31-12-2019T23%3A50&compare_start_date=01-01-2018T00%3A00&groupby=hour}.
\newblock [Last accessed date: 2021-02-19].

\end{thebibliography}
